\documentclass[]{aa}
\usepackage{psfig}
\usepackage{times}
\usepackage{graphics}

\def\la{\;
\raise0.3ex\hbox{$<$\kern-0.75em\raise-1.1ex\hbox{$\sim$}}\; }
\def\ga{\;
\raise0.3ex\hbox{$>$\kern-0.75em\raise-1.1ex\hbox{$\sim$}}\; }

\newcommand{\zabs}{$z_{\rm abs}\,$}
\newcommand{\zem}{$z_{\rm em}\,$}

\newcommand{\kms}{km~s$^{-1}\,$}
\newcommand{\cm}{cm$^{-2}\,$}
\newcommand{\cmm}{cm$^{-3}\,$}

\begin{document}

\title{Spectral shape of 
the UV ionizing background and  \ion{He}{ii} absorption 
at redshifts $1.8 < z < 2.9$\thanks{Based on observations obtained at the VLT Kueyen telescope 
(ESO), the W. M. Keck Observatory and the HST}
}

\author{I. I. Agafonova\inst{1}\thanks{On leave from Ioffe
Physico-Technical Institute, St. Petersburg, Russia}
\and S. A. Levshakov\inst{1}\thanks{On leave from Ioffe
Physico-Technical Institute, St. Petersburg, Russia}
\and D. Reimers\inst{1}
\and C. Fechner\inst{1}
\and D. Tytler\inst{2}
\and R. A. Simcoe\inst{3}
\and A. Songaila\inst{4}
}

\institute{Hamburger Sternwarte, Universit\"at Hamburg,
Gojenbergsweg 112, D-21029 Hamburg, Germany
\and Department of Physics and Center for Astrophysics and Space Sciences,
University of California, San Diego, MS 0424,
La Jolla, CA 92093-0424, USA
\and MIT Center for Space Research, 77 Massachusetts Ave, 37-664B,
Cambridge, MA 02139, USA
\and Institute of Astronomy, University of Hawaii, 2680 Woodlawn Drive,
Honolulu, HI 96822, USA 
}

\offprints{S. A. Levshakov  \protect \\lev@astro.ioffe.rssi.ru}
\date{received date; accepted date}

\abstract{}
{The shape of the UV ionizing background is reconstructed from
optically thin metal absorption-line systems identified in spectra of
\object{HE 2347--4342}, \object{Q 1157+3143}, and \object{HS 1700+6416}
in the redshift interval $1.8 < z < 2.9$.
}
{The systems are analyzed by means of the Monte Carlo Inversion method
completed with the spectral shape recovering procedure.}
{The UVB spectral shape fluctuates at $2.4 < z < 2.9$ mostly due to
radiative transfer processes in the clumpy IGM. At $z \la 1.8$,
the IGM becomes almost transparent both in the \ion{H}{i} and
\ion{He}{ii} Lyman continua and the variability of the spectral
shape comes from diversity of spectral indices describing the 
QSO/AGN intrinsic radiation.
At $z > 2.4$, the recovered spectral shapes show intensity depression
between 3 and 4 Ryd due to \ion{He}{ii} Ly$\alpha$ absorption in
the IGM clouds (line blanketing) and continuous medium (true Gunn-Petersen
effect). The mean \ion{He}{ii} Ly$\alpha$ opacity estimated from
the depth of this depression corresponds within 1-2$\sigma$ to the
values directly measured from the \ion{H}{i}/\ion{He}{ii} Ly$\alpha$
forest towards the quasars studied.
The observed scatter in $\eta = N$(\ion{He}{ii})/$N$(\ion{H}{i}) and
anti-correlation between $N$(\ion{H}{i}) and $\eta$ can be explained by
the combined action of variable spectral softness and differences
in the mean gas density between the absorbing clouds.
Neither of the recovered spectral shapes show features which can be
attributed to the putative input of radiation from soft sources
like starburst galaxies.   
}
{}

\authorrunning{I. I. Agafonova et al.}
\titlerunning{ UVB and \ion{He}{ii} absorption at $1.8 < z < 3.0$ }
\maketitle

\section{Introduction}

This paper continues our study of the UV ionizing background (UVB) of
the metagalactic radiation in the range 1-10 Ryd as reconstructed from  
optically thin QSO metal absorption systems.
In the previous papers, our analysis of the absorbers at $z \approx 3$ 
(Agafonova et al., 2005, hereafter Paper~I) and in the range
$1.46 < z < 1.8$ 
(Reimers et al., 2006, hereafter Paper~II) has led to the following
conclusions:

\begin{enumerate}
\item[1.]
At  $z \approx 3$, the spectral shape of the UV ionizing radiation 
shows a sharp reduction in flux in
the energy range 3~Ryd $< E <$ 4~Ryd which can be interpreted as 
manifestation of the
\ion{He}{ii} Gunn-Petersen effect~--- \ion{He}{ii} Ly$\alpha$ 
continuous absorption in the intergalactic medium.
\item[2.]
At  $z < 1.8$, the ionizing spectra turn out to be much 
harder at $E \geq 3$~Ryd
as compared to model spectra of Haardt \& Madau (1996, hereafter HM). 
This indicates that intergalactic clouds with $N$(\ion{H}{i}) $> 10^{15}$ \cm 
responsible for the \ion{He}{ii} break
(absorption in the \ion{He}{ii} Lyman continuum) become
more rare at low $z$ than was 
supposed by the cloud statistics in HM.
\item[3.] 
The ionizing background
both at $z \approx 3$ and at $z < 1.8$ is dominated
by radiation from QSO/AGNs. The input of the soft radiation from 
galaxies/stars is negligible, limiting the escape fraction of
the galactic UV photons to $f_{\rm esc} < 0.05$.
\end{enumerate}

\noindent
In the present paper, we reconstruct the underlying  UVB
using metal absorbers from the range $1.8 < z < 2.94$
identified in spectra of three bright QSOs: 
\object{HE 2347--4342} (\zem = 2.88), \object{Q 1157+3143}
(\zem = 2.96), and \object{HS 1700+6416} (\zem = 2.74). 
In the considered redshift range, the shape of the UVB
is closely related to the process of \ion{He}{ii} reionization
in course of cosmic time.
Recent observations of 
the \ion{H}{i}/\ion{He}{ii} Ly$\alpha$ forest towards 
\object{HE 2347--4342} (Shull et al. 2004; Zheng et al. 2004, hereafter
Z04) and \object{HS 1700+6416} (Fechner et al. 2006a, hereafter F06)
revealed ($i$) significant fluctuations of the ratio 
$\eta = N$(\ion{He}{ii})/$N$(\ion{H}{i}),
($ii$) anti-correlation between $\eta$ and $N$(\ion{H}{i}),
and ($iii$) decrease of median $\eta$ towards lower $z$. 
Although a large part of the scatter of $\eta$ seems to be artificial and
caused by noise in \ion{He}{ii} data, errors in \ion{H}{i} and \ion{He}{ii}
column densities, unidentified blends etc. (see F06),
similarity of the results obtained both for the \object{HE 2347--4342}
and \object{HS 1700+6416} sightlines
supposes that spatial fluctuations of $\eta$
do really exist.
These fluctuations suggest a variable softness $S = J_{912}/J_{228}$ 
of the metagalactic radiation field
with higher $\eta$ value corresponding to softer UVB.
A diversity of the spectral indices describing QSO continuum radiation,  
an input of radiation from soft sources of type starburst galaxies, and
radiative transfer through clumpy media are the reasons usually
considered for the UVB variability.
Calculations of the UVB spectra accounting for all these effects
are still well beyond modern computational facilities,
although some attempts to explain the observed scatter of $\eta$
via the modeling of the \ion{H}{i}/\ion{He}{ii} Ly$\alpha$ forest
have been made (Maselli \& Ferrara 2005; Bolton et al. 2006).
Thus, direct reconstructions of the UVB spectral shape from
metal absorption-line systems
at different redshifts and along different 
sightlines can help to clarify the character of the UVB fluctuations 
and to determine the sources of the ionizing background.

The paper is organized as follows.
Sect.~2 contains some qualitative information outlining main steps
in the UVB recovering procedure. 
Selected absorption systems are analyzed 
in Sect.~3.
The obtained results are discussed and summarized in Sects.~4 and ~5.

\section{Recovering the shape of the underlying radiation field}

Computational procedure used to recover the 
spectral shape of the ionizing radiation from 
optically thin metal systems is described in Paper~I.	
Here we briefly outline the basics
needed to understand the results presented below in Sect.~3.

\begin{figure}[t]
\vspace{0.0cm}
\hspace{-0.2cm}\psfig{figure=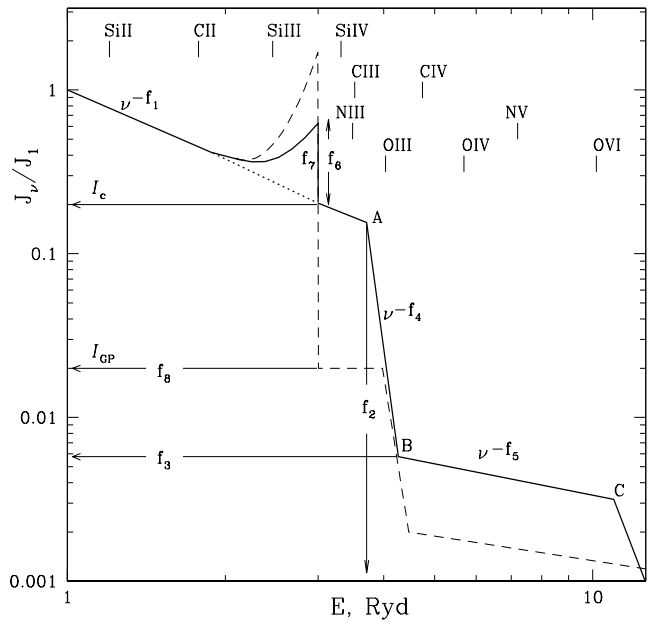,height=10cm,width=9.7cm}
\vspace{-1.5cm}
\caption[]{
Schematic picture of metagalactic 
ionizing spectrum (solid thick line) as predicted by model
calculations (Haardt \& Madau 1996; Fardal et al. 1998).
The spectrum is normalized so that $J_\nu(h\nu =$ 1 Ryd) = 1.
The emission bump at 3 Ryd 
is caused by reemission of \ion{He}{ii} Ly$\alpha$, \ion{He}{ii}
two-photon continuum emission and \ion{He}{ii} Balmer continuum emission
from intergalactic clouds.
The positions of ionization thresholds of different ions are
indicated by tick marks.
The definition of factors $\{f_i\}$ is given in Sect.~2.
Dashed line shows the spectral shape recovered from metal absorbers at
$z \sim 3$.
The continuum depression between 3 and 4 Ryd  ($I_{\rm c} - I_{\rm GP}$)
is due to \ion{He}{ii} Ly$\alpha$ absorption in the IGM   
(the \ion{He}{ii} Gunn-Peterson effect).
}
\label{fg_1}
\end{figure}

Absorption systems are analyzed by means of
the Monte Carlo Inversion (MCI) procedure (Levshakov, Agafonova \& Kegel
2000, hereafter LAK; Levshakov et al. 2002, 2003a,b).
The MCI is based on the assumption that all lines observed in the absorption 
system are formed in a cloud where the gas density and velocity
are random functions of the coordinate $x$ along the cross line. 
It is also assumed that the metal abundance throughout the cloud is constant,
the gas is in thermal and ionization equilibrium 
and optically thin for the ionizing UV radiation.
This implies that ion fractions
are determined entirely by the gas density $n_{\rm H}(x)$
[or, equivalently, by the ionization parameter $U(x) \propto 1/n_{\rm H}$]
and by the spectral energy distribution 
of the background ionizing radiation in the range
1--10 Ryd (defined by the ionization edges of the observed metal ions~--- from
\ion{Si}{ii} to \ion{O}{vi}, see Fig.~\ref{fg_1}).
 
Since the gas density varies from point to point along the sightline, 
the ion fractions vary as well. Thus, for a given point within the
line profile the observed intensity results from a mixture (superposition)
of different ionization states
of the gas due to irregular Doppler shifts of the
local absorption coefficient (see Fig.~1 in LAK). 
The dependence of the ion fraction on
the gas density (ionization curve) is different for each ion
and this is why the line profiles differ in spite of being produced 
by the same gas cloud.

The following physical
parameters are directly estimated in the computational procedure:
the mean ionization parameter $U_0$,
the total hydrogen column density $N_{\rm H}$,
the line-of-sight velocity dispersion, $\sigma_{\rm v}$, and
density dispersion, $\sigma_{\rm y}$, of the bulk material
[$y \equiv n_{\rm H}(x)/n_0$],
and the chemical abundances $Z_{\rm a}$ of all elements
included in the analysis.
With these parameters we can further calculate
the column densities for different species $N_{\rm a}$, and
the mean kinetic temperature $T_{\rm kin}$. 
Given the absolute intensity of the UV background, the mean
gas number density, $n_0$, and the line-of-sight thickness, $L$, 
of the absorber can be evaluated as well.

\begin{figure}[t]
\vspace{0.0cm}
\hspace{-0.2cm}\psfig{figure=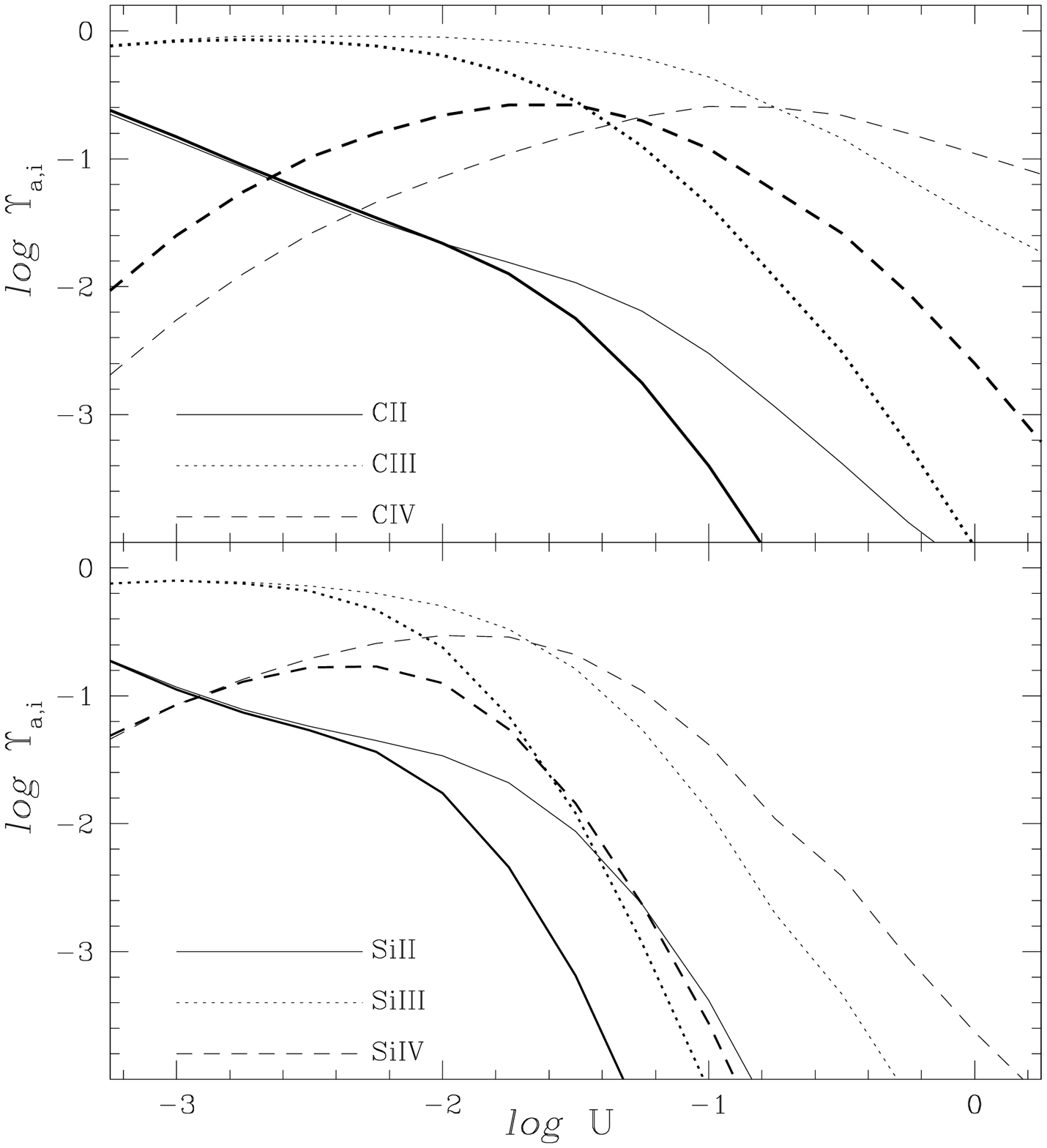,height=10cm,width=9.7cm}
\vspace{-0.5cm}
\caption[]{
Ionization fractions of carbon ({\it upper panel}) and silicon 
({\it lower panel}) calculated for metallicity $Z = 0.01Z_\odot$. 
Thick curves correspond to the UBV shown by the solid line 
in Fig.~\ref{fg_1}, thin curves~--- to the UVB shown by dashed line.
}
\label{fg_2}
\end{figure}

An advanced version of the MCI completed with the
procedure of the spectral shape recovering (referred to as MCISS hereafter)
is iterative and works as follows. 
At first, some basic ionizing spectrum (e.g. HM for corresponding redshift) 
is taken as an initial approximation. 
To calculate the ionization curves needed to carry out the MCI calculations
this basic UVB is inserted into the photoionization code CLOUDY (Ferland 1997)
together with some guesses about relative element abundances and metallicity.
Then the MCI run is performed with the ionization curves taken 
as external parameters. 
If the resulting (recovered) relative abundances and the gas metallicity differ 
from the initial guess, 
the ionization curves are re-calculated with these current quantities
and new MCI runs are performed till  
the recovered metallicity and element ratios correspond to those used 
to calculate the ionization curves. 
If this solution reproduces all line profiles within their uncertainties
and no other peculiarities arise (like relative overabundances of
elements which are well beyond observational and theoretical constraints), 
then we can consider the assumed ionizing spectrum as consistent with 
observations. If, however, the line intensities come
over-/underestimated and/or relative element ratios are odd, then the
spectral shape of the ionizing radiation is to be adjusted.

\begin{figure*}[t]
\vspace{0.0cm}
\hspace{-0.2cm}\psfig{figure=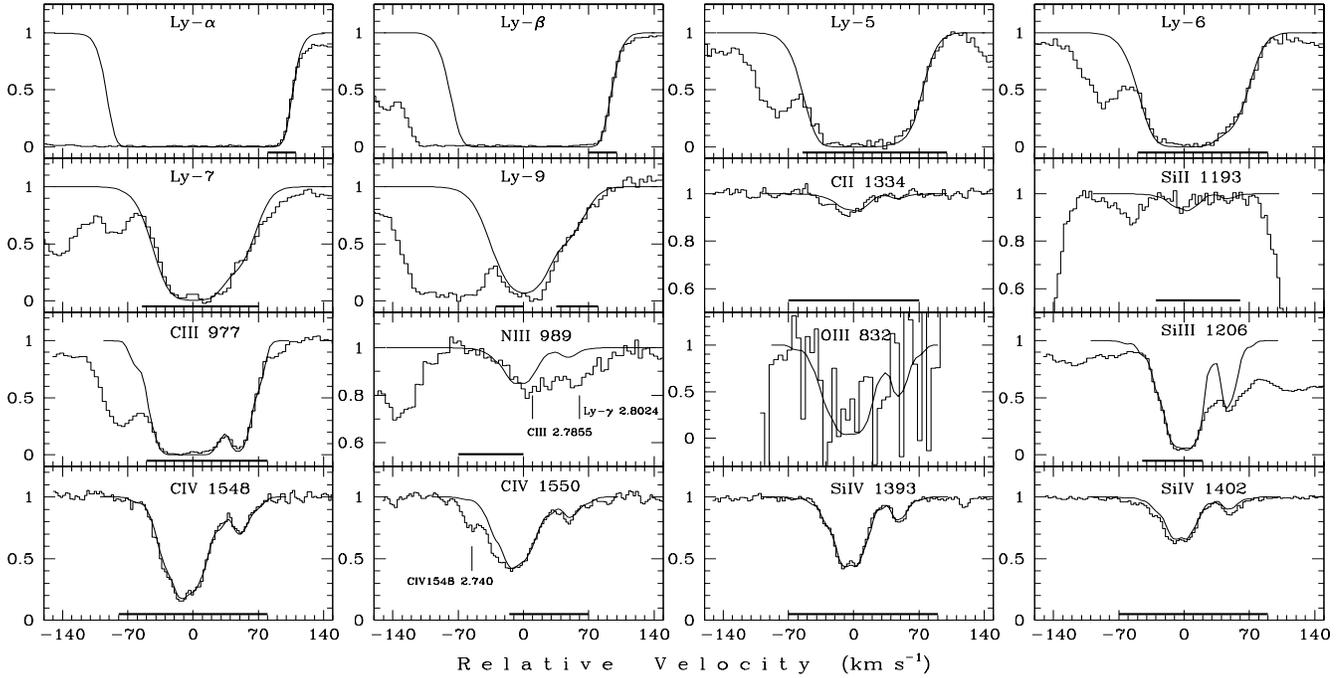,height=16cm,width=18cm}
\vspace{-6.5cm}
\caption[]{
Hydrogen and metal absorption lines associated with the \zabs = 2.735 system
towards \object{HE 2347--4342} (histograms).
Synthetic profiles corresponding to the
recovered ionizing spectrum (dashed line in Fig.~\ref{fg_4})
are plotted by the smooth curves. 
The physical parameters are listed 
in Table~\ref{tb_1}, Col.~2.
Bold horizontal lines mark pixels included in the 
objective function of the MCI procedure.
The zero radial velocity is fixed at $z = 2.73576$.
The central positions of blends are indicated by tick marks.
}
\label{fg_3}
\end{figure*}

The spectral shape adjustment procedure  itself
is developed on base of the response surface methodology
used in the theory of experimental design. It includes
$(i)$ the parameterization of the spectral shape by means of
a set of variables (called `factors'), $(ii)$ the choice of a
quantitative measure (called `response') to evaluate goodness
of a trial spectral shape, and $(iii)$ the estimation of a direction
in the factor space which leads to a spectrum with the better goodness.
Moving along this direction, we come from the initial spectral shape to 
the one with
better characteristics concerning the fitting of the observed line intensities.
Now the MCI calculations are carried out with this newly obtained 
ionizing spectrum and the whole procedure is
repeated till the optimal spectral shape is found, i.e. one 
which allows to reproduce the observed intensities of all lines
without any physical inconsistencies.

How well the spectral shape can be recovered depends on the number of
metal lines involved in the analysis: the more lines of different
ionic transitions of different elements are detected in the absorption
system, the better constrained is the allowable set of the spectral shapes.
Any available a priori information concerning, e.g., the 
expected element ratios  should be also
taken into consideration in order to distinguish between possible solutions.

Fig.~\ref{fg_1} schematically  shows
the spectrum of the metagalactic radiation as
obtained in model calculations (Haardt \& Madau 1996; Fardal et al. 1998) 
and illustrates its parameterization by a set of factors.
These are:
$f_1$~-- the slope (power law exponent) between 1 Ryd and the
\ion{He}{ii} ionization break; 
$f_2$~-- the energy at which the \ion{He}{ii} break starts
(point $A$);
$f_3$~-- the depth of the \ion{He}{ii} ionization break,
$\log (J_{\rm B}/J_{\rm A})$;
$f_4$~-- the slope  between $A$ and $B$;
$f_5$~-- the slope  between $B$ and $C$;
$f_6, f_7$~-- the height and width of the
\ion{He}{ii} re-emission bump.
The energy of the far UV cut off (point $C$) when taken above 100 Ryd 
does not effect the fractional ionizations of ions we are interesting in. 
In all computations described in the subsequent sections
this energy and the slope after it are kept fixed at 
128 Ryd and $-1.5$, respectively.

Studying the UVB at $z \approx 3$, we found
that metagalactic ionizing spectra
reveal a sharp depression in the intensity between 3 and 4 Ryd which 
was interpreted as manifestation of the \ion{He}{ii} Gunn-Petersen (GP) 
effect\footnote{In the present context we consider the GP effect
as being caused by combined action of line blanketing and \ion{He}{ii}
Ly$\alpha$ absorption in continuous intergalactic medium (true GP
effect).} (Paper~I).
Thus, additional factor $f_8$ was introduced to describe the depth of the GP
depression approximated by a straight step.
Knowing this depth, $I_{\rm GP}$, and the continuum level 
$I_{\rm c}$ at $E = 3$ Ryd (see Fig.~\ref{fg_1}), the effective \ion{He}{ii}
Ly$\alpha$ line opacity due to absorption in the intergalactic 
diffuse gas can be estimated 
as $\tau_{\rm eff} = \ln(I_{\rm c}/I_{\rm GP})$.

Several absorption systems in the present work 
have neutral hydrogen column densities in the range
$10^{16}$~\cm  $ < N$(\ion{H}{i}) $< 10^{17}$ \cm, and thus 
are optically thin in the hydrogen Lyman continuum, but 
can be optically thick in the \ion{He}{ii} Lyman continuum. 
In Paper~I we found that such systems can be described with an average 
ionizing spectrum which reflects
both the external (\ion{He}{ii} GP absorption) 
and local effects (strong attenuation at $E > 4$ Ryd due to 
\ion{He}{ii} continuum absorption in the cloud itself and an enhanced peak 
at $E = 3$ Ryd due to recombining emission of
\ion{He}{ii} Ly$\alpha$).
An example of such average spectrum is plotted by the dashed line in 
Fig.~\ref{fg_1} .

Fig.~\ref{fg_2} shows ionization curves of carbon and silicon ions
calculated for this spectrum in comparison to those calculated for
the spectrum of HM ($z = 3$).
The intensity depletion in the range $3 < E < 4$ Ryd 
strongly affects 
ionization fractions of carbon leading to enhanced ratios of
\ion{C}{ii}/\ion{C}{iv} and \ion{C}{iii}/\ion{C}{iv}  
at $\log U \ga -2.5$. The dependence of silicon 
ion fractions on $I_{\rm GP}$ is rather weak, but they are
sensitive to photons with energies $E \leq 3$ Ryd~---
an enhanced intensity in this range leads to a higher   
 \ion{Si}{iv}/\ion{Si}{ii} ratio. 

To complete this section, some words should be said about the evaluation of 
$\eta$. 
In the systems under study, we observe lines of neutral hydrogen, but
we do not have spectral ranges with the corresponding \ion{He}{ii} lines. 
Nevertheless,
assuming that all lines arise in the same gas and reconstructing the velocity
and density distributions of the absorbing material from available hydrogen 
and metal line profiles, we are able to estimate the column density of 
single ionized helium as well.
The ionization corrections of \ion{He}{ii} needed for this estimation 
are calculated using  CLOUDY with the
recovered UVB spectral shape, metallicity and element abundance
ratios as inputs.

\section{Analysis of individual absorption systems}

From a wealth of metal absorption systems identified in the spectra of
\object{HE 2347--4342},  \object{Q 1157+3143}, and \object{HS 1700+6416} 
only a few turned out to be suitable for
the UVB recovering. These metal absorption-line systems 
are described in detail in the subsections below.

In general, the uncertainties of the fitting parameters $U_0$, $N_{\rm H}$,
$\sigma_v$, $\sigma_y$, and $Z_a$ listed in Tables 1, 2 and 3
 are about 15\%--20\%
(for data with S/N $\ga 30$).

All computations  were performed with laboratory wavelengths and
oscillator strengths taken from Morton (2003) for $\lambda >$ 912
\AA\, and from Verner et al. (1994) for $\lambda <$ 912 \AA. Solar
abundances were taken from Asplund et al. (2004). 
Note that their solar abundance of nitrogen, 
N/H~= $6\times10^{-5}$ ,
is 1.4 times (0.15 dex) lower then that
previously reported in Holweger (2001).

The intensity 
$J_{912}$ needed to convert the obtained $U_0$ values into
the gas density was taken from HM.

\subsection{Absorption systems towards \object{HE 2347--4342} }

\begin{figure}[t]
\vspace{0.0cm}
\hspace{-0.5cm}\psfig{figure=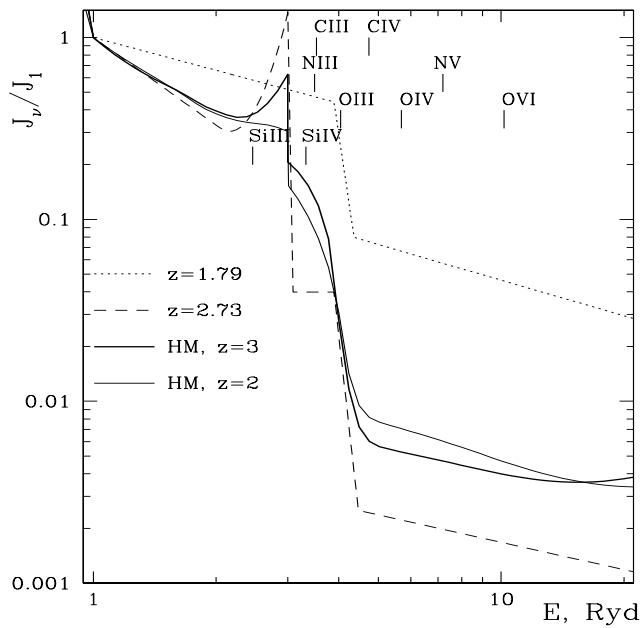,height=10cm,width=12cm}
\vspace{-1.3cm}
\caption[]{
The ionizing background spectrum predicted for $z = 3$ and $z = 2$
in HM (thick and thin solid lines, respectively)
and the spectra restored 
for $z = 2.73$ (dashed lines) and
$z = 1.79$ (dotted line). The normalization is the same as in
Fig.~\ref{fg_1}.
}
\label{fg_4}
\end{figure}

We used for our analysis the VLT/UVES spectrum of  
\object{HE 2347--4342}. Observations and data processing
are described in Shull et al. (2004).

\subsubsection{Absorption systems at $z = 2.735$, $2.739$ and $2.741$ }

These three absorption systems have a joint column density 
of $N$(\ion{H}{i})~= $2.8\times10^{17}$ \cm\, (measured
from Ly$\alpha$ discontinuity in the QSO spectrum). 
The system at \zabs = 2.735 (Fig.~\ref{fg_3}) exhibits
all silicon and carbon transitions from single to triple ionized species, 
\ion{N}{iii} $\lambda989$~\AA\, (partially blended) and 
\ion{O}{iii} $\lambda832$~\AA\, (albeit from very noisy part of the spectrum).
Expected positions of \ion{O}{vi} $\lambda\lambda1032$, 1037~\AA\, 
are blocked with Ly$\alpha$ forest absorptions.
Test calculations with the HM ionizing
spectrum ended up with synthetic line profiles which significantly 
overestimated the observed intensity of \ion{Si}{ii} and
underestimated intensities of \ion{Si}{iv} and \ion{C}{ii}. 
Such behavior of synthetic lines 
along with the procedure of spectral shape adjustment were 
already described in detail in Paper~I, Sect.3.1.
The final spectral shape that we obtained
is shown by the dashed line
in Fig.~\ref{fg_4} with corresponding physical parameters given in 
Table~\ref{tb_1}, Col.2 and synthetic
profiles plotted by the smooth lines in Fig.~\ref{fg_3}. 

The adjusted ionizing spectrum is much softer at $E > 4$ Ryd 
than the initial spectrum of HM
and has a strong emission peak at $E = 3$ Ryd 
(needed to reproduce the observed \ion{Si}{ii}/\ion{Si}{iv} ratio). 
At the same time, the observed ratio of \ion{C}{ii}/\ion{C}{iv} 
cannot be reproduced 
without a significant drop in the intensity  between 3 and 4 Ryd. 
The obtained UVB shape resembles that of \zabs = 2.9171 system towards 
\object{HE 0940--1050} (Paper~I, Sect.3.1) 
and reflects both local (attenuation at $E > 4$ Ryd and enhanced
\ion{He}{ii} Ly$\alpha$ emission at $E = 3$ Ryd)
and external (GP depression at $3 < E < 4$ Ryd due to 
intergalactic \ion{He}{ii} Ly$\alpha$ absorption) effects. 
From the $I_{\rm GP}$ value the effective optical depth of the intergalactic
\ion{He}{ii} Ly$\alpha$ absorption at \zabs = 2.735 is estimated as
$\tau^{\rm He\,II}_{\rm GP} = 1.45\pm0.2$. 
The calculated \ion{He}{ii} column
density is $7.4\times10^{18}$ \cm which gives $\eta = 150$.

\begin{table*}[t]
\centering
\caption{ Physical parameters of the \zabs = 2.735, 2.739,
2.741 and 1.796 metal absorbers towards \object{HE 2347--4342 } 
derived by the MCISS procedure with the recovered UV background spectra
shown in Fig.~\ref{fg_4}. 
Column densities are given in \cm }
\label{tb_1}
\begin{tabular}{lcccc}
\hline
\noalign{\smallskip}
{\footnotesize Parameter}$^a$ & 
{\footnotesize \zabs = 2.735} & 
{\footnotesize \zabs = 2.739} &
{\footnotesize \zabs = 2.741} & {\footnotesize \zabs = 1.796} \\[-2pt]
(1) & (2) & (3) & (4) & (5) \\
\noalign{\smallskip}
\hline
\noalign{\smallskip}
$U_0$& 3.3E--2 & 4.4E--3 & 4.3E--2 & 2.0E--3 \\[-1pt]
$N_{\rm H}$&2.3E20 & 7.3E19 & 5.9E20 & 2.0E17 \\[-1pt]
$\sigma_{\rm v}$, \kms & 25.0 & 24.0 & 21.0 & 15.0 \\[-1pt]
$\sigma_{\rm y}$& 0.6 & 0.5 & 0.7 & 0.6 \\[-1pt]
$Z_{\rm C}$&2.4E--6 & 1.2E--6 & 7.3E--8 & 7.4E--4 \\[-1pt]
$Z_{\rm N}$&$<$2.0E--7 & $\la$8.0E--8 & $\ldots$ & $\ldots$ \\[-1pt]
$Z_{\rm O}$&5.0E--6 & $\ldots$ & $\ldots$ & 1.3E--3 \\[-1pt]
$Z_{\rm Mg}$&$\ldots$ & $\ldots$ & $\ldots$ & (1.4-2.5)E--4 \\[-1pt]
$Z_{\rm Al}$&$\ldots$ & $\ldots$ & $\ldots$ & 1.1E--1 \\[-1pt]
$Z_{\rm Si}$&4.3E--7 & 1.5E--7 & 1.0E--8 & 9.5E--5 \\[-1pt]
$Z_{\rm Fe}$&$\ldots$ & $\ldots$ & $\ldots$ & 2.3E--5 \\[-1pt]
$[Z_{\rm C}]$&$-2.00$ &$-2.30$ & $-3.53$ & 0.48 \\[-1pt]
$[Z_{\rm N}]$&$< -2.5$ & $\la -2.87$ & $\ldots$ &$\ldots$ \\[-1pt]
$[Z_{\rm O}]$&$-2.00$ & $\ldots$ & $\ldots$ & 0.45 \\[-1pt]
$[Z_{\rm Mg}]$&$\ldots$ & $\ldots$ & $\ldots$ & 0.60-0.85 \\[-1pt]
$[Z_{\rm Al}]$&$\ldots$ & $\ldots$ & $\ldots$ & 0.6 \\[-1pt]
$[Z_{\rm Si}]$&$-1.90^c$ & $-2.34^c$ & $-3.53^c$ & 0.46 \\[-1pt]
$[Z_{\rm Fe}]$&$\ldots$ & $\ldots$ & $\ldots$ & $-0.08$ \\[-1pt]
$N$(H\,{\sc i})&$(4.9\pm0.5)$E16 & $(8.6\pm1.0)$E16 & 
$(1.4\pm0.2)$E17 & $(9.1\pm2.0)$E15 \\[-1pt]
$N$(O\,{\sc i}) &$\ldots$ & $\ldots$ & $\ldots$ & $(1.0\pm0.2)$E13 \\[-1pt]
$N$(C\,{\sc ii}) &$(7.1\pm0.5)$E12 & $(4.1\pm0.4)$E12 &  
$(6.2\pm0.6)$E11 & 5.9E13$^b$ \\[-1pt]
$N$(Mg\,{\sc ii}) &$\ldots$ & $\ldots$ & $\ldots$ & (0.9-1.6)E13 \\[-1pt]
$N$(Si\,{\sc ii}) &$(1.6\pm0.2)$E12 & $(5.5\pm0.5)$E11 & 
$(1.0\pm0.2)$E11 & $(1.4\pm0.1)$E13 \\[-1pt]
$N$(Fe\,{\sc ii}) &$\ldots$ & $\ldots$ & $\ldots$ & $(9.4\pm0.5)$E11 \\[-1pt]
$N$(C\,{\sc iii}) &$(3.7\pm0.4)$E14 & 6.9E13$^b$ & 3.0E13$^b$ & $\ldots$ \\[-1pt]
$N$(N\,{\sc iii}) &$\la 2.5$E13 & $\la 4.6$E12 & $\ldots$ & $\ldots$ \\[-1pt]
$N$(O\,{\sc iii}) &$(7.5\pm2.0)$E14 & $\ldots$ & $\ldots$ & $\ldots$ \\[-1pt]
$N$(Al\,{\sc iii}) &$\ldots$ & $\ldots$ & $\ldots$ & $(7.4\pm0.7)$E11 \\[-1pt]
$N$(Si\,{\sc iii}) &$(2.5\pm0.2)$E13 & 6.9E12$^b$ & 
1.6E12$^b$ & $(4.9\pm1.0)$E12 \\[-1pt]
$N$(C\,{\sc iv}) &$(1.2\pm0.1)$E14 & $(4.4\pm0.5)$E12 & 
$(9.5\pm1.5)$E12 & $(1.2\pm0.1)$E13 \\[-1pt]
$N$(Si\,{\sc iv}) &$(2.0\pm0.2)$E13 & $2.0\pm0.2)$E12 & 
$(1.1\pm0.1)$E12 & $(2.9\pm0.3)$E12 \\[-1pt]
$\langle T \rangle$, K & 2.9E4 & 2.0E4 & 3.2E4 & 5.5E3 \\[-1pt]
$n_{\rm H}$, \cmm & 9.0E--4 & 6.0E--3 & 7.0E--4 & 2.2E--2 \\[-1pt]
$L$, kpc & 90 & 4 & 280 & 0.004 \\
$N$(He\,{\sc ii}) &7.4E18$^b$ & 4.3E18$^b$ & 2.0E19$^b$ & 2.0E16$^b$ \\[-1pt]
$\eta$ & $150\pm20$ & $50\pm5$ & $140\pm20$ & $2\pm2$ \\[-1pt]
$\tau^{\rm He\,II}_{\rm GP}$ (at 3 Ryd) &$1.45\pm0.2$ & & & 0 \\
\noalign{\smallskip}
\hline
\noalign{\smallskip}
\multicolumn{5}{l}{\scriptsize $^aZ_{\rm X} = N_{\rm X}/N_{\rm H}$;
$[Z_{\rm X}] = \log (N_{\rm X}/N_{\rm H}) -
\log (N_{\rm X}/N_{\rm H})_\odot$ }\\[-1pt]
\multicolumn{5}{l}{\scriptsize $^b$calculated using the
velocity--density distributions
estimated from hydrogen and metal profiles}\\[-1pt]
\multicolumn{5}{l}{\scriptsize {$^c$actual abundance may be 0.2 dex higher~---
see the last paragraph in Sect.~3.1.1} }
\end{tabular}
\end{table*}

\begin{figure*}[t]
\vspace{0.0cm}
\hspace{0.1cm}\psfig{figure=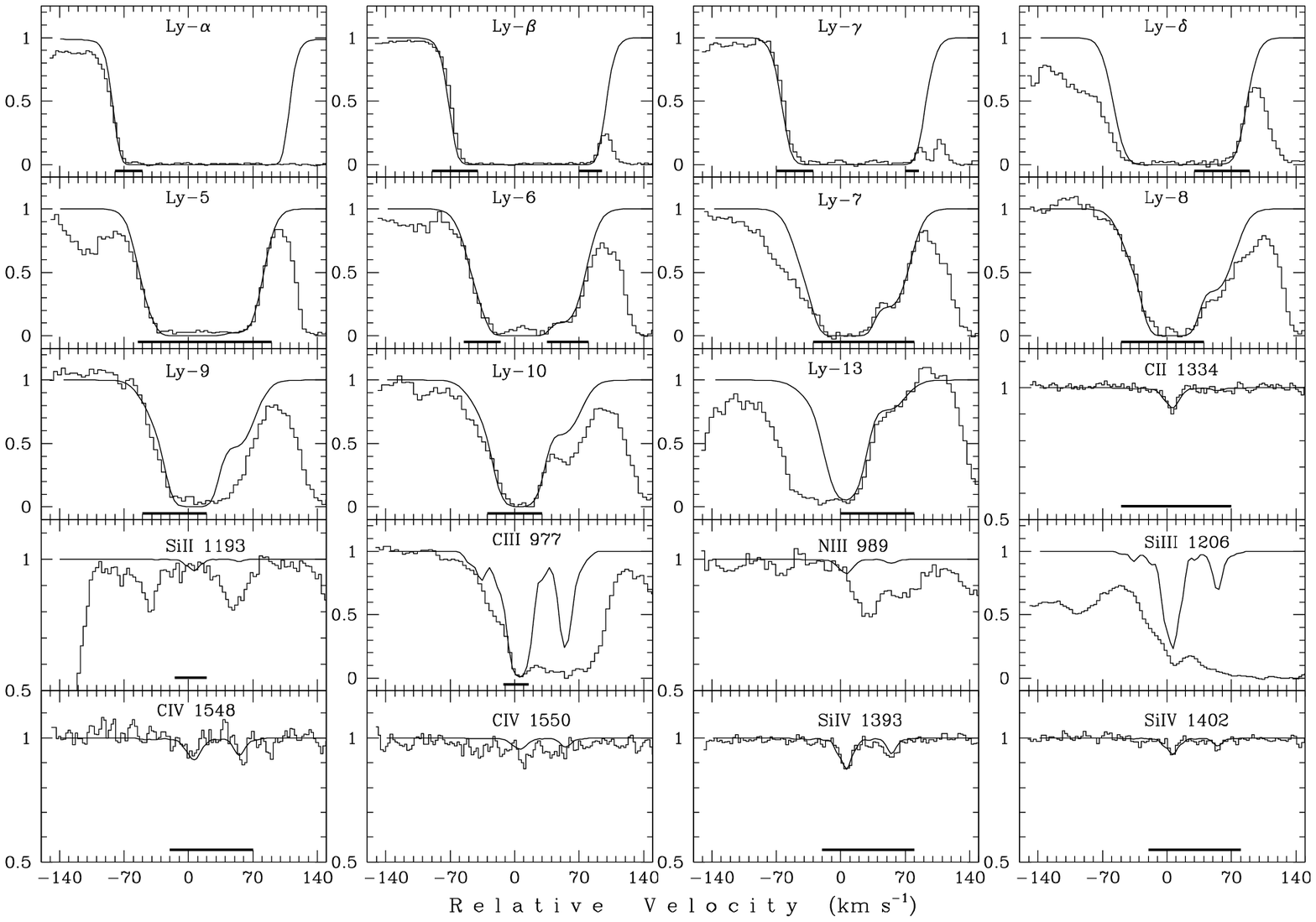,height=13.85cm,width=18cm}
\vspace{-4.6cm}
\caption[]{
Same as Fig.~\ref{fg_3} but for the \zabs = 2.739 system
(\object{HE 2347--4342}).  
The zero radial velocity is fixed at $z = 2.73914$. 
The estimated physical parameters 
are listed in Table~\ref{tb_1}, Col.~3.
}
\label{fg_5}
\end{figure*}
\begin{figure*}[t]
\vspace{0.0cm}
\hspace{0.1cm}\psfig{figure=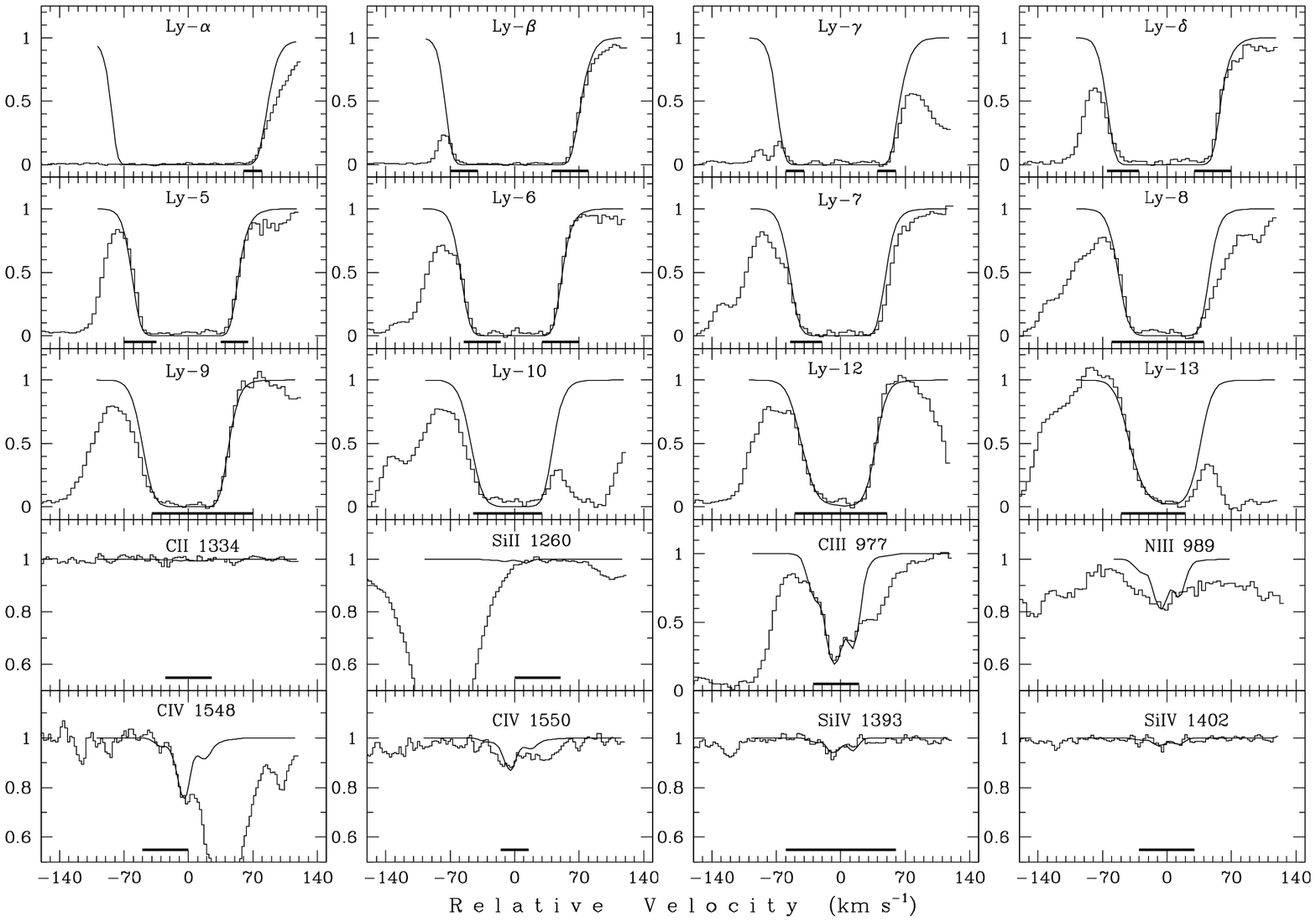,height=13.85cm,width=18cm}
\vspace{-4.6cm}
\caption[]{
Same as Fig.~\ref{fg_3} but for the \zabs = 2.741 system  
(\object{HE 2347--4342}).  
The zero radial velocity is fixed at $z = 2.74097$. 
The estimated physical parameters 
are listed in Table~\ref{tb_1}, Col.~4.
}
\label{fg_6}
\end{figure*}

Two close absorption systems at \zabs = 2.739 and \zabs = 2.741 have metal
lines either heavily blended or weak and noisy leading to the consequence that 
the observed  profiles can be reproduced within
a broad range of spectral shapes. In any case, the UVB recovered from 
the \zabs= 2.735 system is consistent with these systems and 
delivers reasonable physical parameters. 
They are listed in Table~\ref{tb_1}, Cols.~2 and 3, with corresponding synthetic
profiles plotted in Figs.~\ref{fg_5} and \ref{fg_6} (smooth  lines). 
Note an extremely low metallicity of
[$Z$]~$\simeq -3.5$ in the sub-LLS at \zabs = 2.741 
along with its huge linear size of
$L> 250$ kpc. In spite of all uncertainties immanent to this absorber, 
these values can be considered as reliable: the  
\ion{C}{iii}/\ion{C}{iv} ratio is similar to that found 
in the \zabs = 2.735 system (it implies the same mean 
ionization parameter and gas density),
but there is 
three times more neutral hydrogen and an order of 
magnitude less \ion{C}{iii}, \ion{C}{iv} and \ion{Si}{iv}. 
As far as we know, this is the {\it lowest metallicity} ever
directly measured in a Ly$\alpha$ forest absorber.

An important result is that a significant difference in 
$\eta$ between the \zabs = 2.739
absorber ($\eta = 50$) and two other absorbers ($\eta \simeq 150$) stems not from
variations in the ionizing background 
(all quantities have been obtained with the same ionizing spectrum)
but entirely from differences in the gas density: 
$n_{\rm H} = 6.0\times10^{-3}$ \cmm at \zabs = 2.739 
and (7-9)$\times10^{-4}$ \cmm at \zabs = 2.741 and 2.735
(see also Fig.~\ref{fg_19}  below).
 
It is also worth noting that these three absorbers are located in the
vicinity of the \ion{He}{ii} transmission window at 
$2.716 < z < 2.726$ marked by `K' in Figs.~1$f$ and 5 in Z04. 

In all three systems we measured [Si/C]~$\simeq 0-0.1$. In order to
interpret this value correctly the following considerations should be
taken into account. The system at \zabs = 2.735 used to restore the UVB
is optically thin in \ion{H}{i} but opaque in \ion{He}{ii}
which supposes that radiation transfer effects can be significant.
The average spectrum is appropriate to describe the line profiles, but
it can produce biased abundances. Computational experiments with 
different spectral shapes of the UVB have shown that softening at $E > 4$ Ryd
affects more significantly the abundance of silicon making the recovered
ratio [Si/C] lower. In general, the change of [Si/C] does not exceed
0.2 dex. This allows to set an upper bound to [Si/C] in the these
systems: [Si/C]~$\la 0.3$. This value is fully consistent with [Si/C]
measured recently in an extremely metal-poor DLA:
[C]~= $-2.83\pm0.05$, [Si]~= $-2.57\pm0.05$ (Erni et al. 2006), and with
measurements in metal-poor stars:
[Si/C]~$< 0.3$ at [Mg]~$< -3$ (Cayrel et al. 2004). 

The systems show a strong underabundance of nitrogen compared 
to the solar value: [N/C]~$< -0.5$
 
\begin{figure*}[t]
\vspace{0.0cm}
\hspace{0.1cm}\psfig{figure=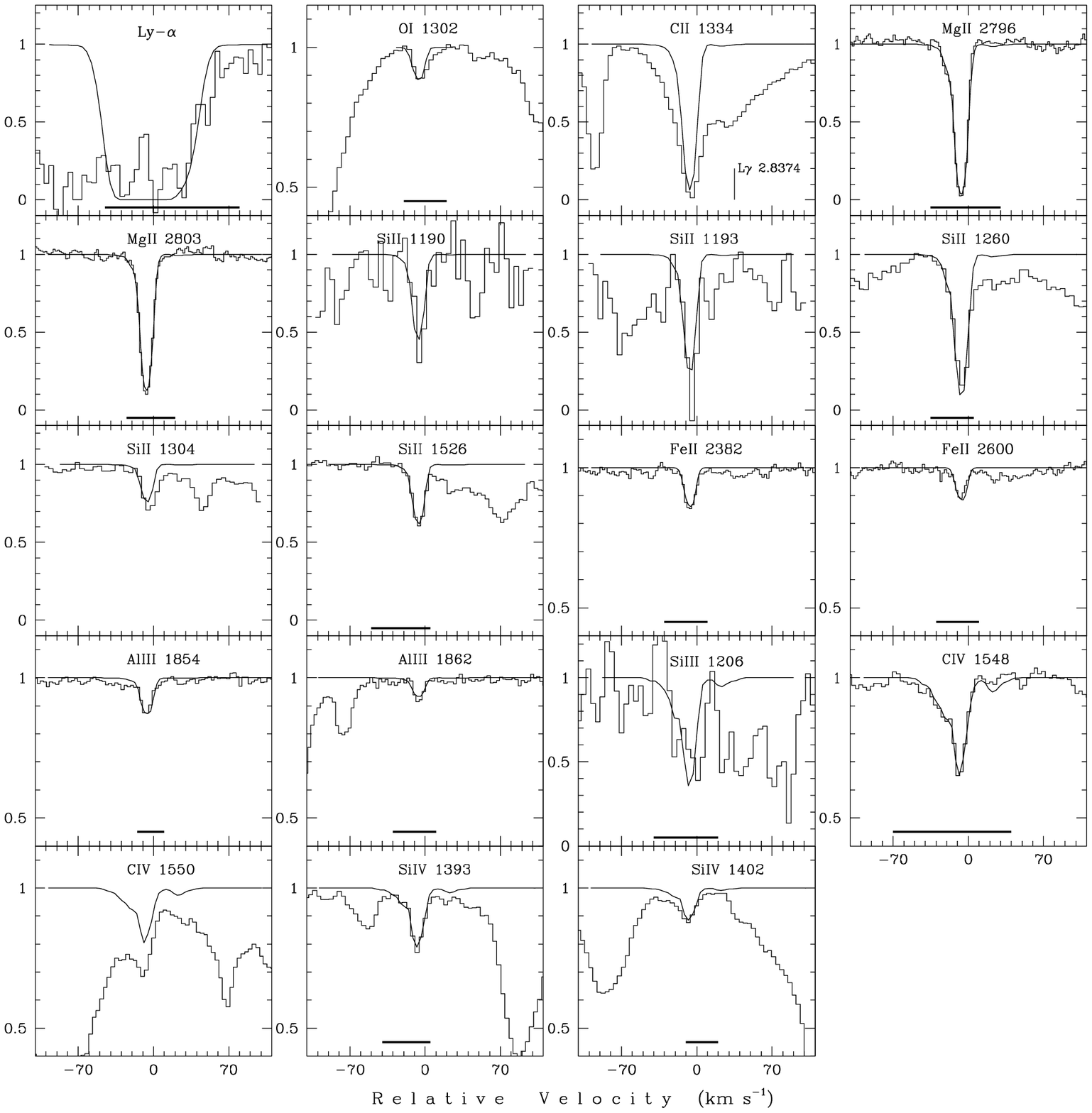,height=16.0cm,width=18cm}
\vspace{-0.6cm}
\caption[]{
Same as Fig.~\ref{fg_3} but for the \zabs = 1.796 system  
(\object{HE 2347--4342}).  
Synthetic profiles (smooth curves) were calculated with the
recovered ionizing spectrum shown by the dotted line in Fig.~\ref{fg_4}.
The zero radial velocity is fixed at $z = 1.7963$. 
The estimated physical parameters 
are listed in Table~\ref{tb_1}, Col.~5.
Synthetic profiles of \ion{Mg}{ii} are calculated for the abundance
$2.5\times10^{-4}$ (see text).
Blends are indicated by tick marks.
}
\label{fg_7}
\end{figure*}

\begin{figure*}[t]
\vspace{0.0cm}
\hspace{-3.0cm}\psfig{figure=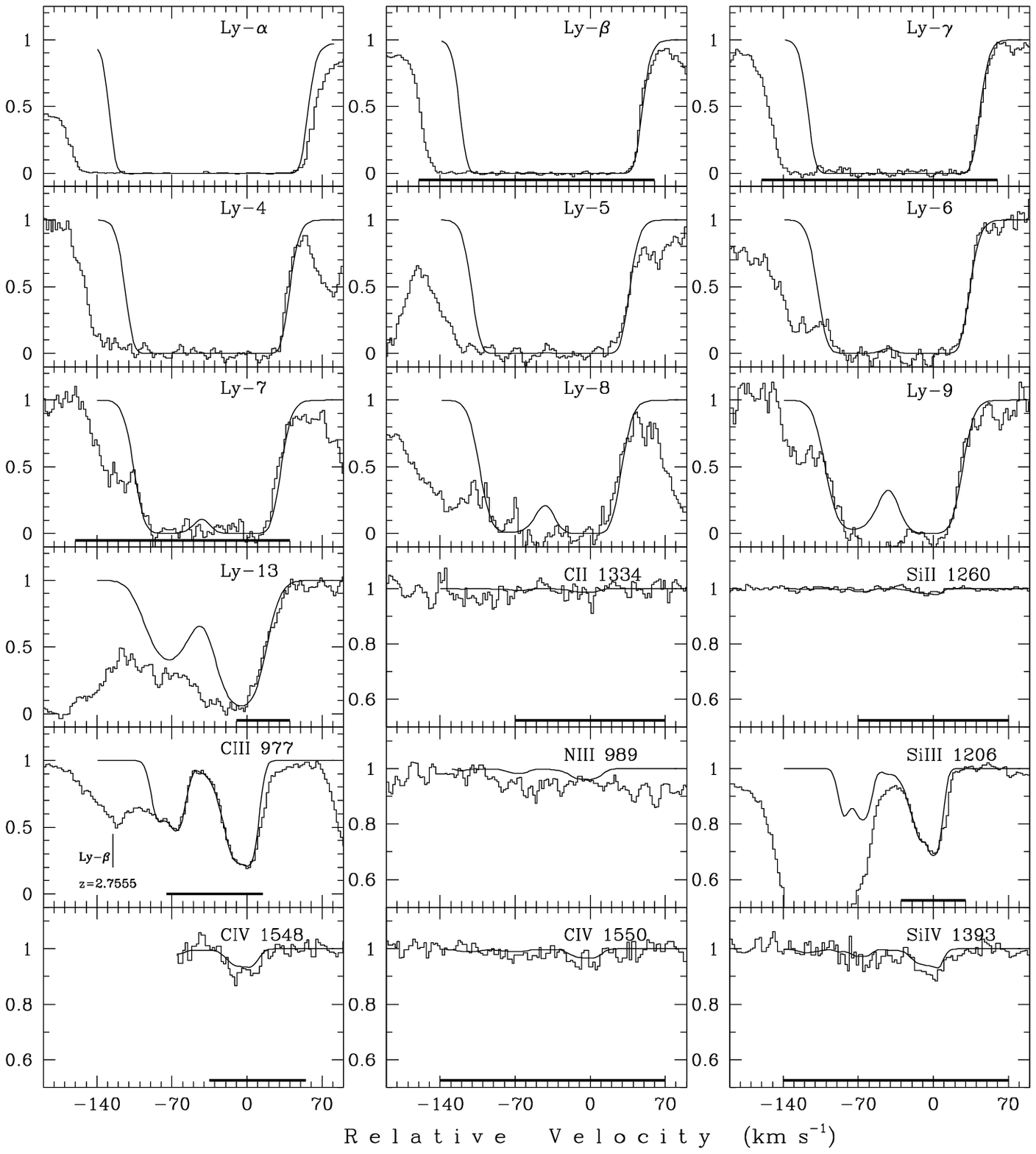,height=16.0cm,width=22cm}
\vspace{-3.0cm}
\caption[]{
Same as Fig.~\ref{fg_3} but for the \zabs = 2.944 system  towards
\object{Q 1157+3143}.  
Synthetic profiles (smooth curves) were calculated with the
recovered ionizing spectrum shown by the dotted line in Fig.~\ref{fg_9}.
The zero radial velocity is fixed at $z = 2.9444$. 
The estimated physical parameters 
are listed in Table~\ref{tb_2}, Col.~2.
Blends are indicated by tick marks.
}
\label{fg_8}
\end{figure*}

\subsubsection{Absorption system at \zabs = 1.7963}

This system was firstly identified through its unusually strong 
\ion{Mg}{ii} lines (Fig.~\ref{fg_7}). 
It also contains absorption lines of \ion{O}{i} $\lambda1302$ \AA, 
\ion{C}{ii} $\lambda1334$ \AA\ (blended), 
\ion{Si}{ii} $\lambda\lambda1260$, 1193, 1190, 1526, and 1304 \AA, 
\ion{Fe}{ii} $\lambda\lambda2600$, 2382 \AA, 
\ion{Al}{iii} $\lambda\lambda1854$, 1862 \AA, 
\ion{Si}{iii} $\lambda1206$ \AA, 
\ion{C}{iv} $\lambda1548$ \AA, 
and \ion{Si}{iv} $\lambda\lambda1393$, 1402 \AA.\, 
Unfortunately, positions of \ion{N}{v} $\lambda\lambda1238$, 1242 \AA\
coincide with Ly$\alpha$ forest absorptions, 
and both lines of the \ion{O}{vi} doublet
fall beyond the available wavelength coverage.
From hydrogen lines, only a noisy Ly$\alpha$ is present and its blue wing 
is blended. 
However, in spite of this uncertainty the available set of metal lines 
and especially the silicon transitions make
it possible to estimate the shape of the underlying UVB ionizing radiation. 

Firstly, the 
\ion{O}{i} $\lambda1302$ \AA\ line is observed together with \ion{C}{iv}. 
\ion{O}{i} traces neutral hydrogen and its ion fraction is practically 
independent on the UVB spectral shape at $E > 1$ Ryd. 
The presence of \ion{O}{i} supposes the ionization parameter 
$U \leq 10^{-3}$, otherwise the fraction of
neutral oxygen becomes too small and 
the resulting overabundance of oxygen to silicon or carbon 
would exceed 1 dex compared to solar values. 

Secondly, all low ionization metal lines
exhibit simple profiles and have the same FWHM of 16 \kms 
($b$-parameter equals 8.7 \kms), which
assumes turbulent motion as the main source of the line broadening. 
Even with this underestimated
$b$-parameter (thermal component in the broadening of the hydrogen line  
is neglected) the neutral hydrogen column density of 
a few $10^{16}$ \cm leads already to a
significant overestimation of the intensity in the unblended red wing. 
This means that the absorption system is optically thin in the hydrogen 
continuum and the self-shielding as a reason for
the simultaneous presence of \ion{O}{i} and \ion{C}{iv} can be excluded. 
Thus, the only remaining option to observe a pronounced 
\ion{C}{iv} absorption at $U \la 10^{-3}$ is
to have an ionizing spectrum which is very hard at $E > 4$ Ryd.

Actually, such type of spectrum was recovered by us for the absorption 
system at \zabs = 1.78 towards \object{HE 0141--3932}  exhibiting a similar 
set of ions (Reimers et al. 2005a). So, it was naturally to take that
spectral shape as an initial guess for the system under study. 
Test runs have shown that the observed intensities of silicon lines 
can be matched only with metallicities of 2.5-3$Z_\odot$, 
otherwise the \ion{Si}{iii} line comes strongly overestimated for all
tried spectral shapes. 
This behavior is explained by a strong dependence of the silicon
ion fractions on temperature which at metallicities near 
and above solar overrides the
dependence on spectral shape.

These considerations allow us to construct the UVB spectrum 
appropriate for the \zabs = 1.796 absorption system 
(Fig.~\ref{fg_4}, dotted line).
Due to many uncertainties stated above
this spectral shape  
is not unique. 
For example, the power index for the slope at $1 < E < 4$ Ryd 
can be higher or lower. This will mostly affect the ratio C/Si
making it above solar in case of a steeper slope. From this
point of view a hard slope is preferable.
However, the depth of the \ion{He}{ii} break at 4 Ryd can be constrained: 
intensities at $E = 4$ Ryd below 0.06$J_{912}$ 
lead to a strong overabundance of oxygen to silicon and to 
carbon ([O/Si,C]~$> 0.2$), and ought to be excluded.

Physical parameters estimated with the UVB from Fig.~\ref{fg_4} are given in
Table~\ref{tb_1}, Col.~5. 
Strong relative underabundance of iron (three times solar values for
carbon, oxygen and silicon and only 0.8 for iron)
can be considered as reliable since it was reproduced for all tried
spectral shapes. Aluminium is slightly overabundant compared to 
C, O and Si (4 times solar), but the accuracy of this estimation is not
high enough due to uncertainties in the recombination coefficients used in
CLOUDY.
As for magnesium, a spread of its abundance (4-7 solar)
is explained by the saturation of \ion{Mg}{ii} $\lambda2796$ \AA\,
and possible contamination of \ion{Mg}{ii} $\lambda2803$ \AA\, with
a telluric line.

\begin{figure}[t]
\vspace{0.0cm}
\hspace{-0.5cm}\psfig{figure=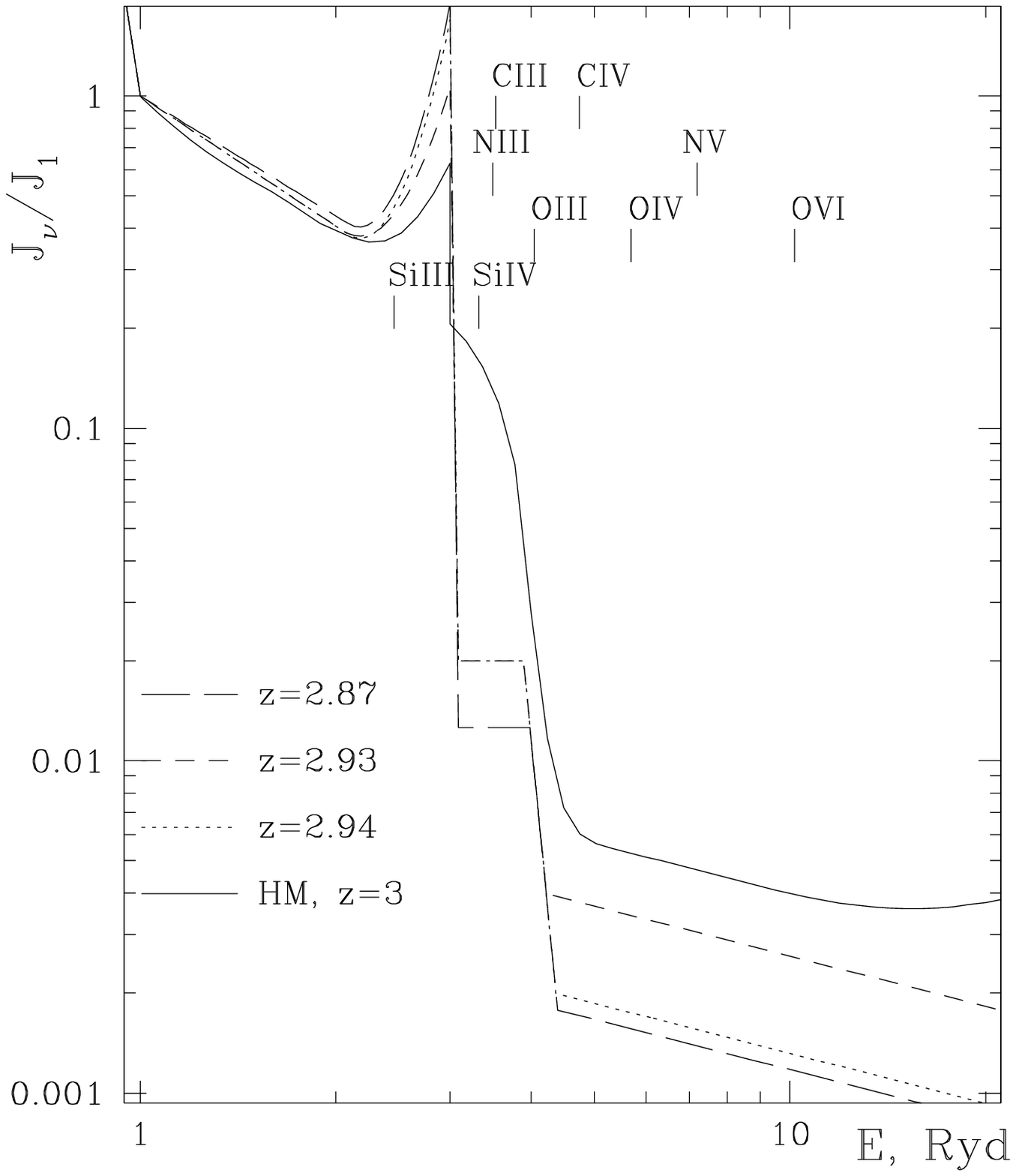,height=9cm,width=12cm}
\vspace{-1.3cm}
\caption[]{
Same as Fig.~\ref{fg_4} but for the systems at \zabs = 2.944,
2.939, and 2.875 towards \object{Q 1157+3143}.
}
\label{fg_9}
\end{figure}

\begin{table*}[t]
\centering
\caption{ Physical parameters of the \zabs = 2.944, 2.939,
and 2.875 metal absorbers towards \object{Q 1157+3143}  
derived by the MCISS procedure with the recovered UV background spectra
shown in Fig.~\ref{fg_9}. 
Column densities are given in \cm }
\label{tb_2}
\begin{tabular}{lccc}
\hline
\noalign{\smallskip}
{\footnotesize Parameter}$^a$ & 
{\footnotesize \zabs = 2.944} & 
{\footnotesize \zabs = 2.939} &
{\footnotesize \zabs = 2.875} \\[-2pt]
(1) & (2) & (3) & (4) \\
\noalign{\smallskip}
\hline
\noalign{\smallskip}
$U_0$& 1.6E--2 & 8.8E--3 & 6.9E--2 \\[-1pt]
$N_{\rm H}$&1.4E20 & 8.3E18 & 2.4E20 \\[-1pt]
$\sigma_{\rm v}$, \kms & 20.0 & 22.5 & 25.0 \\[-1pt]
$\sigma_{\rm y}$& 0.6 & 0.6 & 0.5 \\[-1pt]
$Z_{\rm C}$&3.0E--7 & 2.3E--5 & 1.1E--6 \\[-1pt]
$Z_{\rm N}$&$<$3.0E--8 & $<$1.0E--6 & $\ldots$ \\[-1pt]
$Z_{\rm Si}$&4.7E--8 & 3.7E--6 & 1.6E--7 \\[-1pt]
$[Z_{\rm C}]$&$-2.90$ &$-1.02$ & $-2.36$ \\[-1pt]
$[Z_{\rm N}]$&$< -3.3$ & $< -1.8$ & $\ldots$ \\[-1pt]
$[Z_{\rm Si}]$&$-2.84^d$ & $-0.94$ & $-2.34^d$ \\[-1pt]
$N$(H\,{\sc i})&$(1.1\pm0.2)$E17 & $1.2\pm0.1)$E16 & 
$(1.6\pm0.4)$E16 \\[-1pt]
$N$(C\,{\sc ii}) &$(1.4\pm0.2)$E12 & $(9.0\pm0.5)$E12 & 
$(2.0\pm0.3)$E12 \\[-1pt]
$N$(Si\,{\sc ii}) &$(2.5\pm0.2)$E11 &$(1.5\pm0.2)$E12&$(1.8\pm0.2)$E11 \\[-1pt]
$N$(C\,{\sc iii}) &$(3.7\pm0.6)$E13$^b$ &
$(1.8\pm0.2)$E14 &$(1.9\pm0.2)$E14 \\[-1pt]
$N$(N\,{\sc iii}) &$< 5.8$E12 & $\la 7.5$E12 & $\ldots$ \\[-1pt]
$N$(Si\,{\sc iii}) &$(3.4\pm0.4)$E12$^b$ & 2.0E13$^c$ & 
4.2E12$^c$ \\[-1pt]
$N$(C\,{\sc iv}) &$(3.2\pm0.8)$E12 &$(1.1\pm0.3)$E13 &$(4.0\pm0.4)$E13 \\[-1pt]
$N$(Si\,{\sc iv}) &$(1.7\pm0.3)$E12 &$(8.0\pm0.4)$E12 &$(6.8\pm0.5)$E12 \\[-1pt]
$\langle T \rangle$, K & 2.7E4 & 1.8E4 & 3.7E4 \\[-1pt]
$n_{\rm H}$, \cmm & 2.0E--3 & 3.0E--3 & 4.0E--4 \\[-1pt]
$L$, kpc & 27 & 0.86 & 200 \\
$N$(He\,{\sc ii}) &8.5E18$^c$ & 5.4E17$^c$ & 7.0E18$^c$ \\[-1pt]
$\eta$ & $85\pm10$ & $45\pm5$ & $440\pm40$ \\[-1pt]
$\tau^{\rm He\,II}_{\rm GP}$ (at 3 Ryd) 
&\multicolumn{2}{c}{$2.5\pm0.3$} & 2.5-3 \\
\noalign{\smallskip}
\hline
\noalign{\smallskip}
\multicolumn{4}{l}{\scriptsize $^aZ_{\rm X} = N_{\rm X}/N_{\rm H}$;
$[Z_{\rm X}] = \log (N_{\rm X}/N_{\rm H}) -
\log (N_{\rm X}/N_{\rm H})_\odot$;\,\, $^b$partially blended; }\\[-1pt]
\multicolumn{4}{l}{\scriptsize 
$^c$ calculated using the velocity--density distribution
estimated from hydrogen and metal profiles}\\[-1pt]
\multicolumn{4}{l}{\scriptsize {$^d$actual value may be 0.2 dex higher~---
see the last paragraph in Sect.~3.1.1} }
\end{tabular}
\end{table*}

\subsection{Absorption systems towards \object{Q 1157+3143}}

Absorption spectrum  of this quasar was obtained with Keck/HIRES. 
Details on the data reduction are given in Kirkman \& Tytler (1999). 
Additionally, this bright
quasar was observed with HST/STIS to study \ion{He}{ii} Ly$\alpha$ absorption in
the intergalactic matter (Reimers et al. 2005b).

\subsubsection{Absorption system at \zabs = 2.944}

This absorption system (Fig.~\ref{fg_8}) 
lies very close to the quasar (\zem = 2.97, Ly$\alpha$ forest starts
at $z = 2.999$) and could be associated. 
However, \object{Q 1157+3143} does not show proximity
effect. On the contrary, the line density in its vicinity 
is unusually high (Ganguly et al. 2001). 
A set of ions detected in the \zabs = 2.944 system  
is also typical for an intergalactic absorber: associated systems 
are usually highly ionized and show pronounced lines of 
\ion{O}{vi}, \ion{N}{v}, and \ion{C}{iv} along with weak 
\ion{C}{iii}, \ion{Si}{iii}, and \ion{S}{iv}, whereas the system under 
study reveals \ion{C}{iii} and \ion{Si}{iii} lines stronger than 
\ion{C}{iv}, \ion{Si}{iv}, and no \ion{N}{v} absorption.
At the expected position of the \ion{O}{vi} $\lambda1031$ \AA\, line 
only a weak absorption is seen (its identification
as \ion{O}{vi} is ambiguous since \ion{O}{vi} $\lambda1037$ \AA\, is blended).

From the \ion{H}{i} continuum absorption in the observed spectrum 
the neutral hydrogen column density of this
system (together with its neighbor at \zabs = 2.939) is estimated
as $(1.5\pm0.2)\times10^{17}$ \cm. 
Such a high hydrogen content together with rather weak intensities of
the metal lines point to a very low metallicity of the absorbing gas 
which is another argument against the associated nature
of the \zabs = 2.944 system.

Test calculations with the HM ionizing spectrum confirmed 
the metallicity of $0.001Z_\odot$. 
However, with this spectrum the calculated intensity of the 
\ion{C}{iii} line in the range $-70 < v < 70$ \kms\ was 
significantly less than observed. 
Since the position of this line is in the Ly$\alpha$ forest,
possible blends should be taken into account. We do not find any metal
absorption contaminating the \ion{C}{iii} line.
The only candidate for blending 
could be a hydrogen absorption at \zabs = 2.17. 
If we assume that observed profile is mainly caused by
this hydrogen absorption, then  
the kinetic temperature of the gas estimated from the 
line width is $T_{\rm kin} \leq 32 000$~K. This
implies that the ionization parameter $U$ is less than 0.02 
(for metallicity $[Z] < -1.5$, since there are no metal lines detected
at \zabs = 2.17) 
and the linear size of the putative \ion{H}{i} absorber
does not exceed 0.5 kpc. Such small cloud would hardly
survive in the IGM. Another argument in favor of \ion{C}{iii}
is the comparison with the \ion{Si}{iii} line: both lines have
similar widths ($FWHM = 40$ \kms and 36.5 \kms) and the same 
asymmetry. 
Thus, we can assume with high probability
that the absorption at $-70 < v < 70$ \kms\ 
is indeed caused by \ion{C}{iii} from the \zabs = 2.944 system.  
In this case only a spectrum 
with a significant
flux depression between 3 and 4 Ryd makes it possible to fit adequately all 
observed lines. 

The reconstructed spectral shape of the UVB 
is shown in Fig.~\ref{fg_9} by the dotted line, 
the corresponding physical parameters are given in Table~\ref{tb_2}, 
Col.~2, and the synthetic profiles are plotted by the solid lines in 
Fig.~\ref{fg_8}. 

Here we find again a spectrum similar to that from the \zabs = 2.9171 system 
(\object{HE 0940--1050}, Paper~I):
its shape reflects both the external (GP depression) and local effects 
(an enhanced emission peak at 3 Ryd and a sharp break at 4 Ryd)
arising from the reprocession of the incident 
ionizing radiation by the absorbing cloud itself. 

The \ion{He}{ii} Ly$\alpha$ opacity is estimated as 
$\tau^{\rm He\,II}_{\rm GP} = 2.5\pm0.2$. 
The predicted value for the \ion{He}{ii} column
density is $8.5\times10^{18}$ \cm\ which gives $\eta = 85$.

The ratio [Si/C] = 0.06, but as mentioned in Sect.~3.1.1 the actual
value may be 0.2 dex higher: [Si/C]~$ \la 0.26$.

A significant relative underabundance of nitrogen, 
[N/C]~$ < -0.3$, is also worth noting for this system.

\begin{figure*}[t]
\vspace{0.0cm}
\hspace{-2.5cm}\psfig{figure=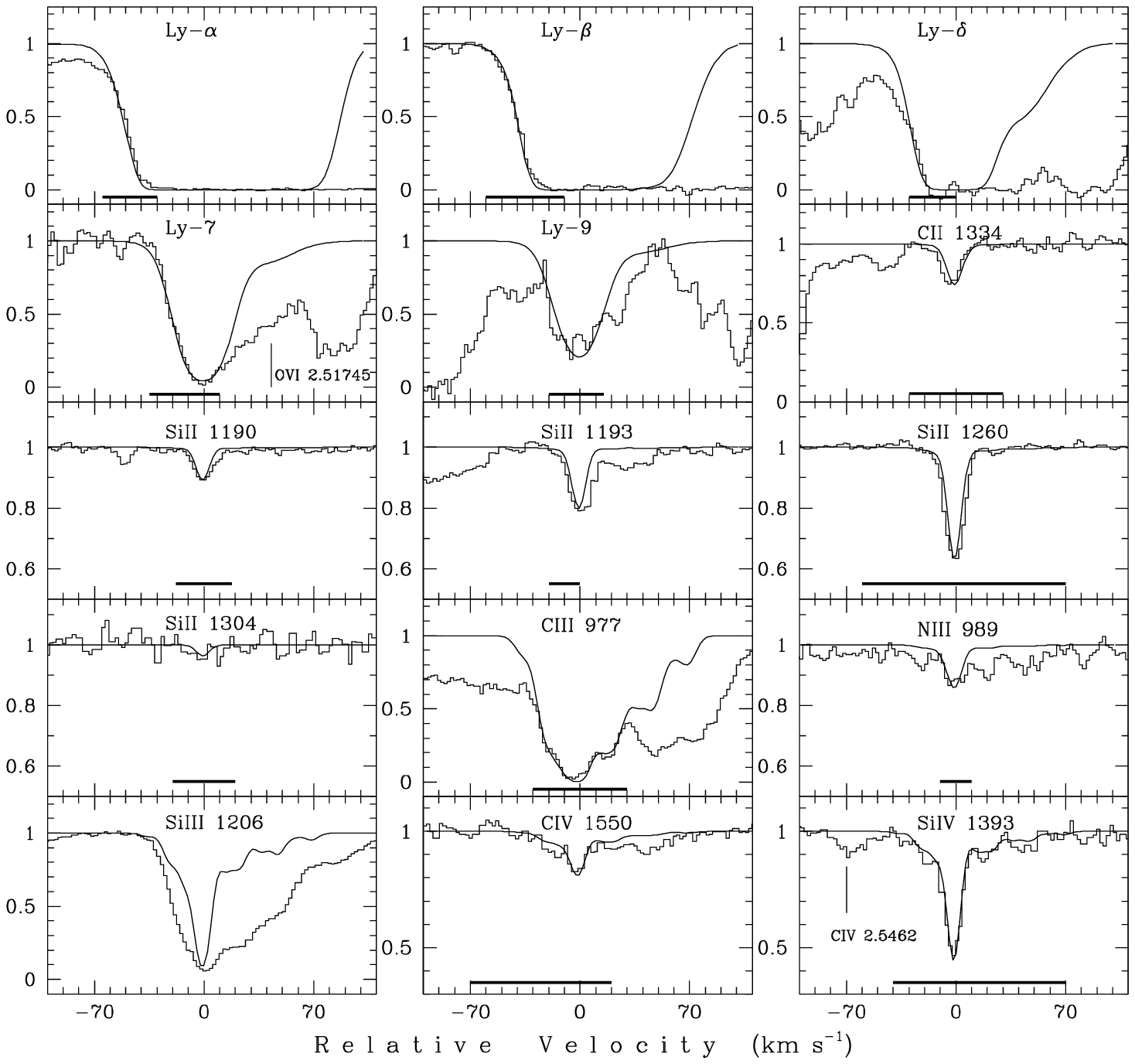,height=16.0cm,width=22cm}
\vspace{-5.0cm}
\caption[]{
Same as Fig.~\ref{fg_3} but for the \zabs = 2.939 system  towards
\object{Q 1157+3143}.  
Synthetic profiles (smooth curves) were calculated with the
recovered ionizing spectrum shown by the short-dashed line in Fig.~\ref{fg_9}.
The zero radial velocity is fixed at $z = 2.93987$. 
The estimated physical parameters 
are listed in Table~\ref{tb_2}, Col.~3.
Blends are indicated by tick marks.
}
\label{fg_10}
\end{figure*}

\subsubsection{Absorption system at \zabs = 2.939}

This system presents metal lines of both low 
(\ion{C}{ii}, \ion{Si}{ii}) and high (\ion{C}{iv}, \ion{Si}{iv})
ionization stages (Fig.~\ref{fg_10}). 
Lines \ion{O}{vi} $\lambda\lambda1031, 1037$ \AA\ 
are blended with strong Ly$\alpha$ forest absorptions.
The spectrum of HM turned out to be inconsistent with the
observed line intensities, the adjusted spectral shape 
is shown in Fig.~\ref{fg_9} by the short-dashed line. 
The corresponding physical parameters are given in Table~\ref{tb_2}, 
Col.~3, and  the synthetic
profiles are plotted by the solid lines in Fig.~\ref{fg_10}.

Although shifted by only $\Delta z = 0.0045$ ($\Delta v = 344$ \kms) 
from the \zabs = 2.944 system, the present absorber 
reveals significantly different physical conditions: 
metallicity two order of magnitude higher
along with a lower ionization parameter (higher gas density) 
and a very small (for an intergalactic absorber)
linear size below 1 kpc.
 
The \zabs = 2.939 system has optical depth $\simeq$ 1 in the 
 \ion{He}{ii} Lyman continuum.
Thus, the attenuation of the incident radiation is probably not
very pronounced and the recovered spectral shape  
can represent 
in this case the external UVB. 
It is seen from Fig.~\ref{fg_9} that  spectrum is softer at 
$E > 4$ Ryd than the model spectrum of HM and has a more pronounced
\ion{He}{ii} Ly$\alpha$ emission peak at $E = 3$ Ryd. 
This is probably a consequence of a high number density of clouds
registered at $z \la 3$ along the sight line towards 
\object{Q 1157+3143}. 

The predicted value of $\eta = 45\pm5$ is
two times lower than $\eta$ in the \zabs = 2.944 system 
and reflects both the higher gas density and 
the ionizing background which is harder at $E > 4$ Ryd 
(see Fig.~\ref{fg_9}).

Noteworthy is again a very strong relative underabundance of nitrogen, 
[N/C]~$< -0.8$.

\begin{figure*}[t]
\vspace{0.0cm}
\hspace{0.2cm}\psfig{figure=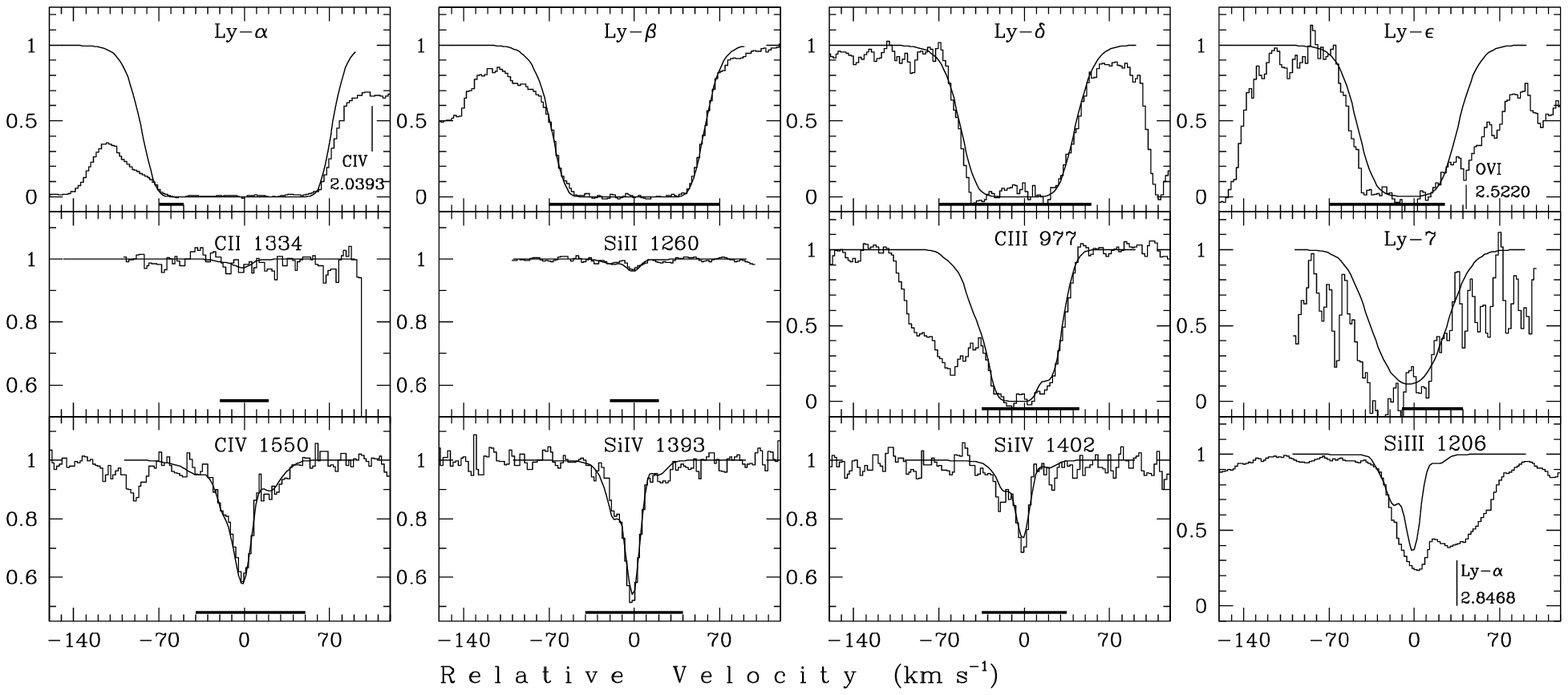,height=14.0cm,width=18cm}
\vspace{-8.0cm}
\caption[]{
Same as Fig.~\ref{fg_3} but for the \zabs = 2.875 system  towards
\object{Q 1157+3143}.  
Synthetic profiles (smooth curves) were calculated with the
recovered ionizing spectrum shown by the long-dashed line in Fig.~\ref{fg_9}.
The zero radial velocity is fixed at $z = 2.87567$. 
The estimated physical parameters 
are listed in Table~\ref{tb_2}, Col.~4.
Blends are indicated by tick marks.
}
\label{fg_11}
\end{figure*}

\begin{figure*}[t]
\vspace{0.0cm}
\hspace{0.2cm}\psfig{figure=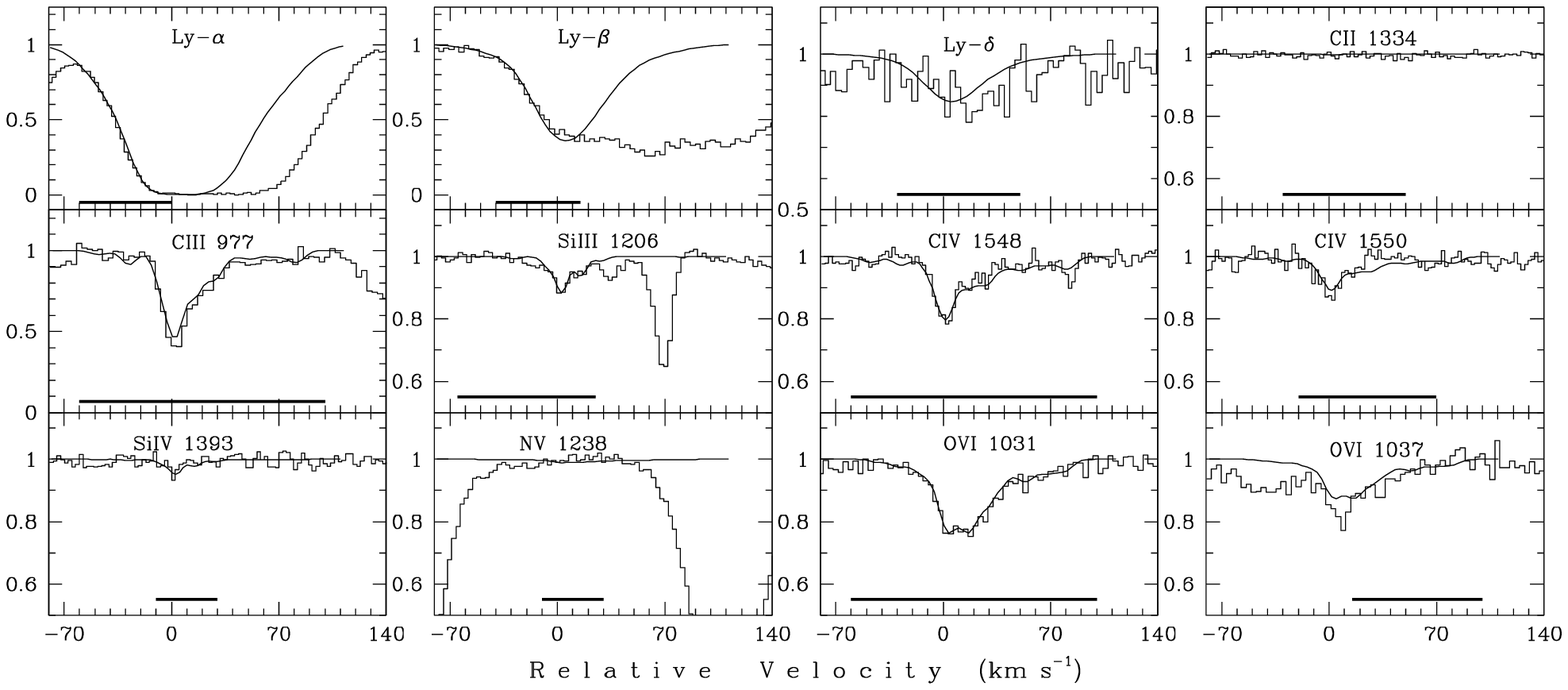,height=14.0cm,width=18cm}
\vspace{-8.0cm}
\caption[]{
Same as Fig.~\ref{fg_3} but for the \zabs = 2.568 system  towards
\object{HS 1700+6416}.  
Synthetic profiles (smooth curves) were calculated with the
recovered ionizing spectrum shown by the dotted line in Fig.~\ref{fg_13}.
The zero radial velocity is fixed at $z = 2.56826$. 
The estimated physical parameters 
are listed in Table~\ref{tb_3}, Col.~2.
}
\label{fg_12}
\end{figure*}

\subsubsection{Absorption system at \zabs = 2.875}

Hydrogen and metal lines identified in this system are shown in 
Fig.~\ref{fg_11}. The \ion{C}{iv} $\lambda1548$ \AA\ line falls in a gap
between echelle orders,
\ion{N}{iii} $\lambda 989$ \AA\ and 
\ion{O}{vi} $\lambda\lambda1031, 1037$ \AA\ are severely blended. 
Note an extremely weak \ion{Si}{ii} $\lambda1260$ \AA\ line
which is nevertheless very well detected 
owing to high S/N~= 150 in this wavelength range.

Test calculations with the HM ionizing spectrum  
showed carbon content [C/H]~$= -2.3$ associated with 
a very high relative overabundance of silicon, 
[Si/C]~$\simeq 0.8$. 
The calculated intensity at the expected position of \ion{C}{iii} came 
significantly underestimated,
but blending cannot be excluded 
(the only candidate~--- \ion{H}{i} Ly$\alpha$ at \zabs = 2.115).
 
The hydrogen column density is not well defined because of high noise 
at the locations of  Ly-7 $\lambda926$ \AA\ and Ly-8 $\lambda923$ \AA,
but, in any case, it does not exceed $3\times10^{16}$ \cm.

The obtained overabundance of silicon points to the ionizing spectrum 
much softer at $E > 4$ Ryd than the HM spectrum that we used.
A chance that high [Si/C] arises from contribution of 
the hypothetical Pop~III stars with masses $M \geq 100M_\odot$
(pair-instability SNe)
seems quite unprobable since such overabundance has never been 
observed even in the extremely metal-poor clouds with 
$Z \la 0.001Z_\odot$ (see Sect.~3.1.1). 
A soft spectrum can be produced either
by the \ion{He}{ii} continuous absorption (both in the IGM and
inside the cloud itself) 
or by the input of radiation 
from soft sources like stars. 
In the present case the first option
is preferable taking into account the observational evidence 
of large amount of \ion{He}{ii} at \zabs~$\simeq 2.87$ 
towards \object{Q 1157+3143} (Reimers et al. 2005b),
and a high value of the ionization parameter $U$
(log(\ion{Si}{iv}/\ion{Si}{ii})~$\simeq 1.6$) which supposes
$\eta > 150$ (see Fig.~19) which assusmes, in turn, that 
the cloud is opaque in \ion{He}{ii}. 

Unfortunately, high noise in the  \ion{C}{ii} $\lambda1334$ \AA\ profile 
and blending of \ion{Si}{iii} hamper the accurate
reconstruction of the ionizing background. 
Calculations with the spectrum from the \zabs = 2.944 system lead to the silicon 
overabundance [Si/C]~$\simeq$ 0.2-0.3. 
Taking into account that the average spectrum gives [Si/C] by $\la 0.2$ dex
lower than the real value, this translates to [Si/C]~$\approx 0.4-0.5$
which is slightly higher than measured elsewhere (Sect.~3.1.1).
A UVB 
with a lowered
by 0.2 dex depression between 3 and 4 Ryd 
(Fig.~\ref{fg_9}, long-dashed line) gives [Si/C]~$\simeq 0$ 
(and reproduces the observed line intensity at the position 
of \ion{C}{iii}).  
These two shapes constrain the acceptable range. 
The physical parameters obtained with the latter spectrum
are listed in Table~\ref{tb_2}, Col.~4. 
The corresponding synthetic profiles are plotted by 
the solid lines in Fig.~\ref{fg_11}.
The calculated column density of 
\ion{He}{ii}, $N$(\ion{He}{ii})~$= 7.0\times10^{18}$ \cm, supposes that the 
cloud is optically thick in the \ion{He}{ii} continuum
and, thus, the restored spectral shape is again a superposition of the
local and external effects.

The mean optical depth of \ion{He}{ii} Ly$\alpha$,  
$\tau^{\rm He\,II}_{\rm GP}$~= 2.5-3, is 
comparable to 
$\tau^{\rm He\,II}_{\rm GP}$~= $2.09\pm0.1$ estimated at \zabs = 2.87
by Reimers et al. (2005b). 

We note the extremely high value of 
$\eta = 440\pm40$ which
is reliably reproduced in all test runs in spite of uncertainties in both 
the neutral hydrogen content and
the spectral shape of the ionizing radiation. 
This value can be explained by the softness
of the effective ionizing radiation field 
and by the low gas density in the absorbing cloud 
(cf. with $\eta = 150$ at
\zabs = 2.944 produced by the ionizing spectrum of similar softness).

\subsection{Absorption systems towards \object{HS 1700+6416} }

This one of the brightest known high redshift quasar was observed 
many times with many instruments, both ground- and
satellite-based. 
Reimers et al. (1992) and
Vogel \& Reimers (1995) firstly reported on 7 strong metal absorption 
systems identified in its 
low-resolution combined optical and UV spectra. 
Later on, Davidsen et al. (1996) measured opacity of 
the intergalactic \ion{He}{ii} Ly$\alpha$ using 
far UV spectrum obtained with the HUT.
Simcoe et al. (2002, 2006) studied physical conditions in the metal 
absorption systems using high-resolution Keck telescope data.
A comprehensive analysis of both \ion{H}{i} and \ion{He}{ii} (FUSE)
spectra was performed by F06 and Fechner \& Reimers (2006). 
For our reconstruction of the underlying UVB spectra we selected  
systems with many different ionic lines, that are in the ionization 
equilibrium and do not show self-shielding effects\footnote{Since
metal systems towards \object{HS 1700+6416}
are well known, the reader may be interested in more details
why other systems were not included in the UVB recovering procedure:
$z = 2.578$~--- two overlapping systems with different metallicities which
cannot be clearly separated; $z = 2.315$~--- probably a non-equilibrial
system with overionized \ion{O}{vi}, may be associated with an AGN/QSO
close to the line of sight; $z = 2.168$~--- the ratios 
\ion{Si}{iv}/\ion{Si}{ii}~$\simeq 40$ and 
\ion{Si}{iv}/\ion{C}{iv}~$\simeq 0.16$
point out to a very soft UVB (cf. data for the $z = 2.875$ system in Table~2),
however, uncertainties in \ion{C}{ii} and \ion{Si}{iii}
(blending ?) and the absence of \ion{C}{iii} $\lambda977$ \AA\
do not allow us
to reconstruct the actual spectral shape;
$z = 1.724$~--- most metal lines are contaminated by the Ly$\alpha$ forest
absorptions. 
}. 
Observational data stem from  the Keck/HIRES 
($\Delta \lambda = 3150-9000$ \AA) and HST/STIS 
($\Delta \lambda = 1155-1700$ \AA) spectra. 
Details on data reduction are given in Simcoe et al. (2002) and
Fechner et al. (2006b).

\begin{figure}[t]
\vspace{0.0cm}
\hspace{-0.5cm}\psfig{figure=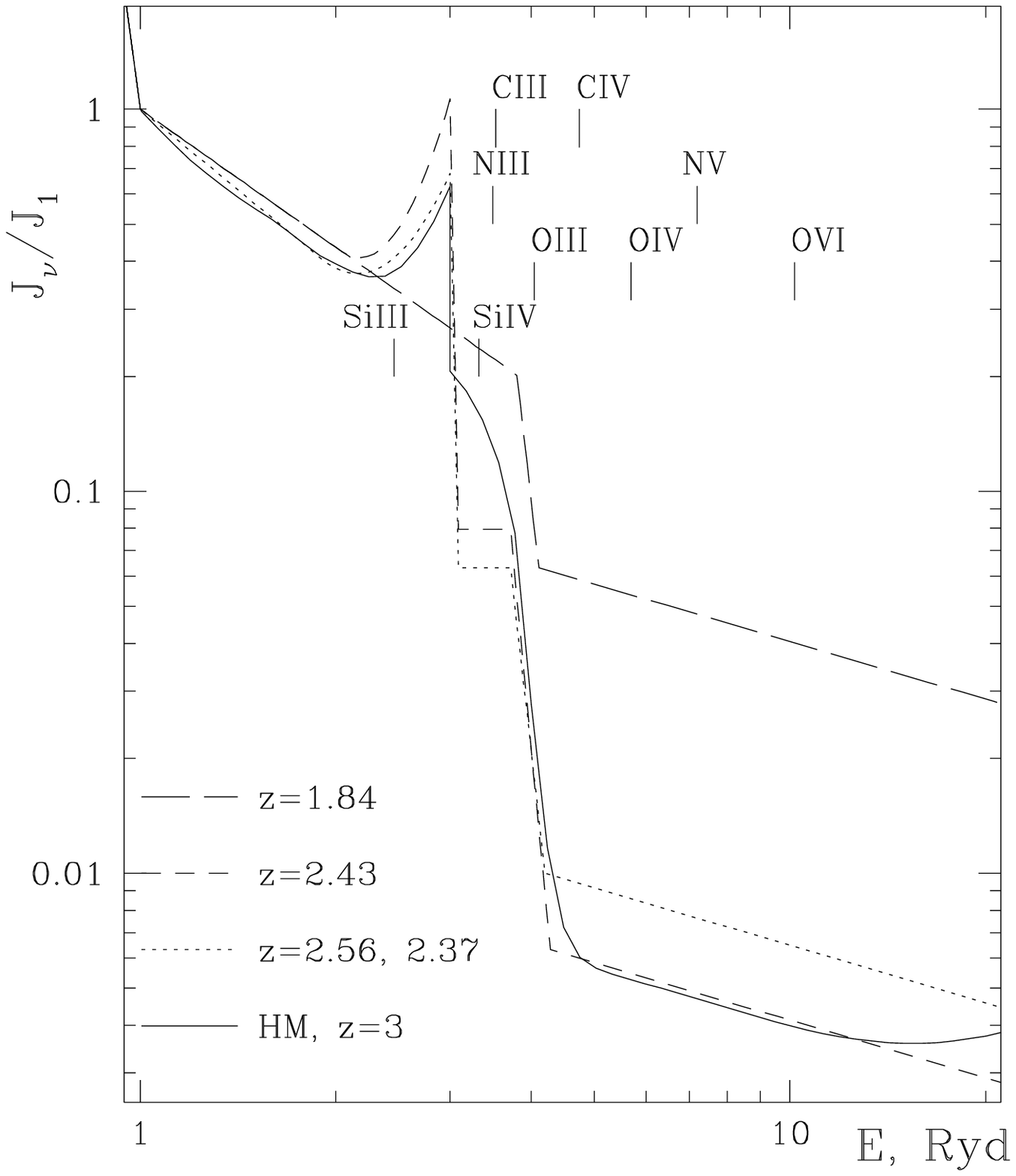,height=9cm,width=12cm}
\vspace{-1.3cm}
\caption[]{
Same as Fig.~\ref{fg_4} but for the systems at \zabs = 2.568,
2.433, 2.438, 2.379, and 1.845 towards \object{HS 1700+6416}.
}
\label{fg_13}
\end{figure}

\subsubsection{Absorption system at \zabs = 2.568}

Hydrogen and metal lines identified in this system are shown in 
Fig.~\ref{fg_12}. Test calculations with the ionizing
spectrum of HM revealed that the red wing of 
\ion{H}{i} Ly$\alpha$ is inconsistent with the assumption
of constant metallicity inside the absorber. 
In the following, the red wings of hydrogen lines are
reconstructed from the density-velocity distributions derived from 
metal lines under the assumption
of constant metallicity.

Calculations with the HM spectrum
produced also strongly underestimated intensity of \ion{C}{iii}.
Blending of \ion{C}{iii} seems rather unprobable since 
the line is weak, narrow and its asymmetric profile is similar to those 
of \ion{C}{iv} and \ion{O}{vi}. 
Thus the shape of the HM ionizing spectrum
was modified.  The result is shown in Fig.~\ref{fg_13} 
by the dotted line. 
The corresponding physical parameters
are given in Table~\ref{tb_3}, Col.~2, and the
synthetic profiles are plotted by the solid lines in Fig.~\ref{fg_12}. 

The mean optical
depth of the intergalactic \ion{He}{ii} Ly$\alpha$ absorption
decreases at redshift \zabs = 2.56 to 
the value $\tau^{\rm He\,II}_{\rm GP} = 1.2\pm0.2$. 
Note again a significant underabundance of nitrogen,
[N/C]~$ < -0.3$.

\begin{table*}[t]
\centering
\caption{ Physical parameters of the \zabs = 2.568, 2.438,
2.433, 2.379, and 1.845 metal absorbers towards \object{HS 1700+6416} 
derived by the MCISS procedure with the recovered UV background spectra
shown in Fig.~\ref{fg_13}. 
Column densities are given in \cm }
\label{tb_3}
\begin{tabular}{lccccc}
\hline
\noalign{\smallskip}
{\footnotesize Parameter}$^a$ & 
{\footnotesize \zabs = 2.568} & 
{\footnotesize \zabs = 2.438} &
{\footnotesize \zabs = 2.433} & 
{\footnotesize \zabs = 2.379} & 
{\footnotesize \zabs = 1.845} \\[-2pt]
(1) & (2) & (3) & (4) & (5) & (6)\\
\noalign{\smallskip}
\hline
\noalign{\smallskip}
$U_0$& 1.3E--1 & 1.8E--2 & 9.4E--3 & 1.1E--1 & 7.9E--3 \\[-1pt]
$N_{\rm H}$&3.3E18 & 1.2E19 & 2.0E19 & 3.2E19 & 2.1E19 \\[-1pt]
$\sigma_{\rm v}$, \kms & 20.5 & 17.0 & 23.0 & 27.0 & 35.0 \\[-1pt]
$\sigma_{\rm y}$& 0.5 & 0.5 & 0.7 & 0.6 & 0.6 \\[-1pt]
$Z_{\rm C}$&2.0E--5 & 9.2E--6 & 1.5E--5 & 2.8E--6 & 8.4E--5 \\[-1pt]
$Z_{\rm N}$&$\la$3.0E--6 & $\la$2.0E--6 & $\ldots$ & $\ldots$ & 1.7E--5 \\[-1pt]
$Z_{\rm O}$&1.0E--4 & $\ldots$ & $\ldots$ & 1.6E--5 & 2.5E--5 \\[-1pt]
$Z_{\rm Mg}$&$\ldots$ & $\ldots$ & $\ldots$ & $\ldots$ & 2.5E--5 \\[-1pt]
$Z_{\rm Al}$&$\ldots$ & $\ldots$ & $\ldots$ & $\ldots$ & 1.0E--6 \\[-1pt]
$Z_{\rm Si}$&6.2E--6 & 1.8E--6 & 2.4E--6 & 8.4E--7 & 1.4E--5 \\[-1pt]
$[Z_{\rm C}]$&$-1.08$ &$-1.42$ & $-1.20$ & $-1.94$ & $-0.46$ \\[-1pt]
$[Z_{\rm N}]$&$\la -1.3$ & $\la -1.5$ & $\ldots$ &$\ldots$ & $-0.55$ \\[-1pt]
$[Z_{\rm O}]$&$-0.65$ & $\ldots$ & $\ldots$ & $-1.52$ & $-0.25$ \\[-1pt]
$[Z_{\rm Mg}]$&$\ldots$ & $\ldots$ & $\ldots$ & $\ldots$ & $-0.25$ \\[-1pt]
$[Z_{\rm Al}]$&$\ldots$ & $\ldots$ & $\ldots$ & $\ldots$ & $-0.36$ \\[-1pt]
$[Z_{\rm Si}]$&$-0.75$ & $-1.24$ & $-1.12$ & $-1.61$ & $-0.36$ \\[-1pt]
$N$(H\,{\sc i})&2.5E14$^b$ & $(3.6\pm1.5)$E15 & 
$(2.6\pm1.0)$E16 & $(2.4\pm0.3)$E15 & 2.5E16$^b$ \\[-1pt]
$N$(C\,{\sc ii}) &$(1.4\pm0.2)$E11 &$(1.3\pm0.3)$E12 &$(1.3\pm0.2)$E13&$\ldots$ 
& 3.6E13$^c$  \\[-1pt]
$N$(Mg\,{\sc ii}) &$\ldots$ & $\ldots$ & $\ldots$ & $\ldots$ & 2.2E12 \\[-1pt]
$N$(Si\,{\sc ii}) &$\ldots$ &$\la 3.0$E11 &$(2.2\pm0.2)$E12 & $\ldots$ 
& $(3.9\pm0.2)$E12 \\[-1pt]
$N$(C\,{\sc iii}) &$(1.2\pm0.2)$E13 &$(6.7\pm0.6)$E13 &$(2.5\pm0.3)$E14 
& $(2.3\pm0.4)$E13 & $\ldots$ \\[-1pt]
$N$(N\,{\sc iii}) &$\ldots$ & $\la 1.4$E13 & $\ldots$&$\ldots$&$\ldots$ \\[-1pt]
$N$(O\,{\sc iii}) &$\ldots$ & $\ldots$ & $\ldots$ & $\ldots$ & 1.6E15$^c$ \\[-1pt]
$N$(Al\,{\sc iii}) &$\ldots$ & $\ldots$ & $\ldots$ & $\ldots$ & 
$(5.8\pm0.6)$E11 \\[-1pt]
$N$(Si\,{\sc iii}) &$(4.2\pm0.3)$1E11 & $(5.0\pm0.3)$E12 & 
$(2.8\pm0.2)$E13  & $(6.7\pm0.3)$E11 & 3.5E13$^d$ \\[-1pt]
$N$(C\,{\sc iv}) &$(1.2\pm0.1)$E13 &$(2.5\pm0.3)$E13 &$(3.6\pm0.2)$E13 
& $(1.6\pm0.1)$E13 & $(4.3\pm0.2)$E14  \\[-1pt]
$N$(O\,{\sc iv}) &$\ldots$ & $\ldots$ & $\ldots$ & $\ldots$ & 1.8E15$^c$ \\[-1pt]
$N$(Si\,{\sc iv}) &$(4.8\pm0.4)$E11 &$(4.0\pm0.2)$E12 &$(1.2\pm0.1)$E13 
& $(5.3\pm0.5)$E11 & $(3.8\pm0.3)$E13 \\[-1pt]
$N$(N\,{\sc v}) &$\ldots$ & $\ldots$ & $\ldots$ & $\ldots$ & 
$(3.5\pm0.3)$E13 \\[-1pt]
$N$(O\,{\sc vi}) &$(3.9\pm0.4)$E13 & $\ldots$ & $\ldots$ & 
$(3.0\pm0.3)$E13 & $\ldots$ \\
$\langle T \rangle$, K & 3.3E4 & 2.6E4 & 2.1E4 & 4.3E4 & 1.5E4 \\[-1pt]
$n_{\rm H}$, \cmm & 2.5E--4 & 1.5E--3 & 3.0E--3 & 3.0E--4 & 4.0E--3 \\[-1pt]
$L$, kpc & 4.0 & 3 & 2 & 38 & 2 \\
$N$(He\,{\sc ii}) &2.6E16$^c$ & 3.7E17$^c$ & 1.1E18$^c$ & 
2.9E17$^c$ & 3.5E17$^c$ \\[-1pt]
$\eta$ & $100\pm10$ & $100\pm10$ & $45\pm5$ & $120\pm20$ & 
$16\pm5$ \\[-1pt]
$\tau^{\rm He\,II}_{\rm GP}$ (at 3 Ryd)
& $1.2\pm0.2$ & \multicolumn{2}{c}{
$1.2\pm0.2$} & $1.2\pm0.2$ & 0 \\
\noalign{\smallskip}
\hline
\noalign{\smallskip}
\multicolumn{6}{l}{\scriptsize $^aZ_{\rm X} = N_{\rm X}/N_{\rm H}$;
$[Z_{\rm X}] = \log (N_{\rm X}/N_{\rm H}) -
\log (N_{\rm X}/N_{\rm H})_\odot$;\,\, $^b$ column
density calculated from the velocity--density }\\[-1pt]
\multicolumn{6}{l}{\scriptsize distribution
estimated from metal profiles assuming constant metallicity within 
the absorber;\,\, $^c$ calculated using the velocity and density}\\[-1pt]
\multicolumn{6}{l}{\scriptsize distributions
derived from hydrogen and metal profiles;\,\,
$^d$ partially blended}
\end{tabular}
\end{table*}

\begin{figure*}[t]
\vspace{0.0cm}
\hspace{0.2cm}\psfig{figure=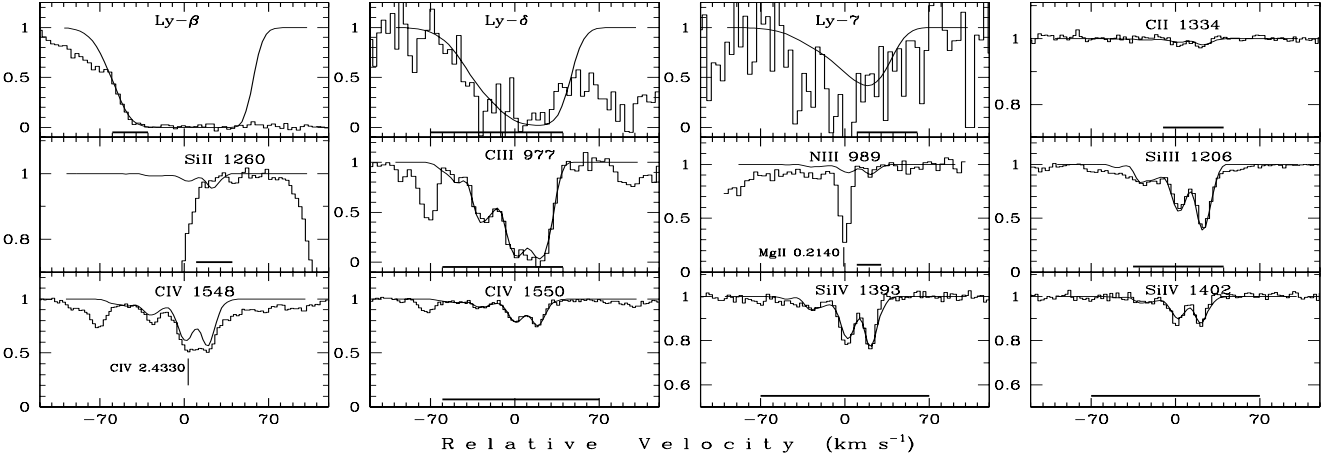,height=14.0cm,width=18cm}
\vspace{-8.0cm}
\caption[]{
Same as Fig.~\ref{fg_3} but for the \zabs = 2.438 system  towards
\object{HS 1700+6416}.  
Synthetic profiles (smooth curves) were calculated with the
recovered ionizing spectrum shown by the short-dashed line in Fig.~\ref{fg_13}.
The zero radial velocity is fixed at $z = 2.4386$. 
The estimated physical parameters 
are listed in Table~\ref{tb_3}, Col.~3.
Blends are indicated by tick marks.
}
\label{fg_14}
\end{figure*}

\begin{figure*}[t]
\vspace{0.0cm}
\hspace{0.2cm}\psfig{figure=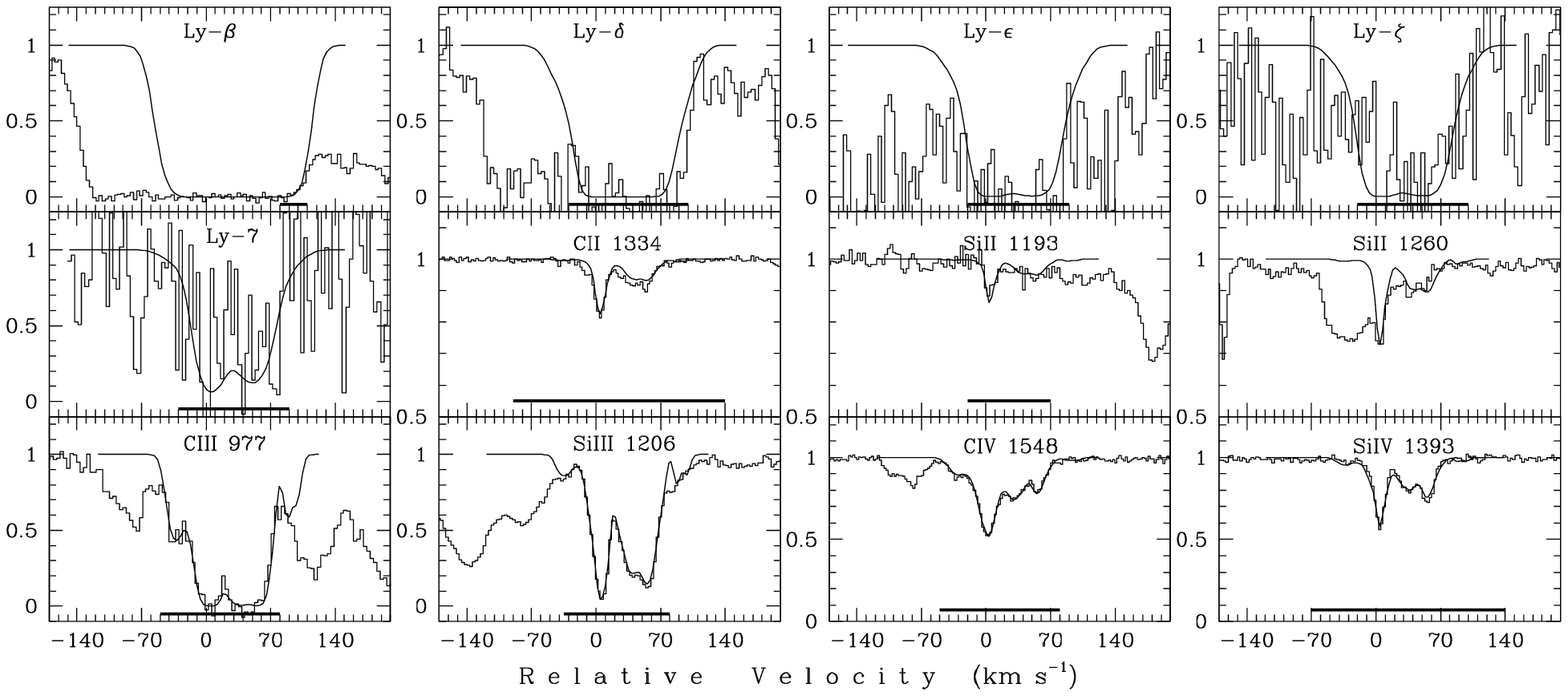,height=14.0cm,width=18cm}
\vspace{-8.0cm}
\caption[]{
Same as Fig.~\ref{fg_3} but for the \zabs = 2.433 system  towards
\object{HS 1700+6416}.  
Synthetic profiles (smooth curves) were calculated with the
recovered ionizing spectrum shown by the short-dashed line in Fig.~\ref{fg_13}.
The zero radial velocity is fixed at $z = 2.4330$. 
The estimated physical parameters 
are listed in Table~\ref{tb_3}, Col.~4.
}
\label{fg_15}
\end{figure*}

\subsubsection{Absorption systems at \zabs = 2.438 and \zabs = 2.433}  

Both absorption systems (Figs.~\ref{fg_14} and \ref{fg_15}) 
are parts of a large absorption complex stretching over 1000 \kms 
(see Fig.~6 in Simcoe et el. 2006) 
and represent dense clumps embedded in a highly ionized gas  
traced by \ion{O}{vi}. 
While this highly ionized gas could be still non-equilibrial, 
the systems under consideration  
have already reached the ionization equilibrium with the background radiation 
owing to their higher gas density and, hence, a shorter
recombination time.
Note that the whole absorbing complex resembles the \zabs = 1.8073 complex
towards \object{HS 0747+4259} described in Sect.~4.1 in Paper~II.

From low-resolution spectrum, the neutral hydrogen column density 
of the whole complex is estimated as
$4.0\times10^{16}$ \cm\ (Vogel \& Reimers, 1995). 
Test calculations with the HM ionizing spectrum resulted in 
an underestimated intensity of 
\ion{C}{ii} $\lambda1334$ \AA\, along with a strong overestimation 
of \ion{Si}{ii} $\lambda1193$ \AA. 
The adjusted spectral shape is shown in Fig.~\ref{fg_13} 
by the short-dashed line. The calculated physical parameters are listed 
in Table~\ref{tb_3}, Cols.~3 and 4, and the
corresponding synthetic profiles are shown by the solid lines in 
Figs.~\ref{fg_14} and \ref{fg_15}. 

Although the spectral shape differs from that recovered 
for the \zabs = 2.568 system, the mean \ion{He}{ii} optical depth
is the same: $\tau^{\rm He\,II}_{\rm GP} = 1.2\pm0.2$. 
This value is to be compared with
$\tau^{\rm He\,II}_{\rm GP} = 1.0\pm0.07$ 
measured at $z = 2.4$ in the low resolution far UV spectrum of
\object{HS 1700+6416} by Davidsen et al. (1996), and 
with $\tau^{\rm He\,II}_{\rm GP} = 0.74\pm0.34$ at $z = 2.45$
measured from FUSE data by F06.  
It should be noted, however, that latter value can be underestimated
due to problems with zero level flux in the FUSE spectrum
(see Sect.~2 in F06).

Another point worth noting is the variation we find 
in $\eta$:  $\eta = 45$ at \zabs = 2.433 and $\eta = 100$ at \zabs = 2.438
(Table~3),
which occurs because of the density differences in the clouds under study.

The value of $\eta$ at \zabs = 2.438 is close to the $\eta$ estimated
for the previous \zabs = 2.568 system although the gas density in these two
absorbers is quite different. This result is a consequence of softer 
UVB at \zabs = 2.438 
(see Fig.~\ref{fg_19} in Sect.~4). 

\begin{figure*}[t]
\vspace{0.0cm}
\hspace{0.2cm}\psfig{figure=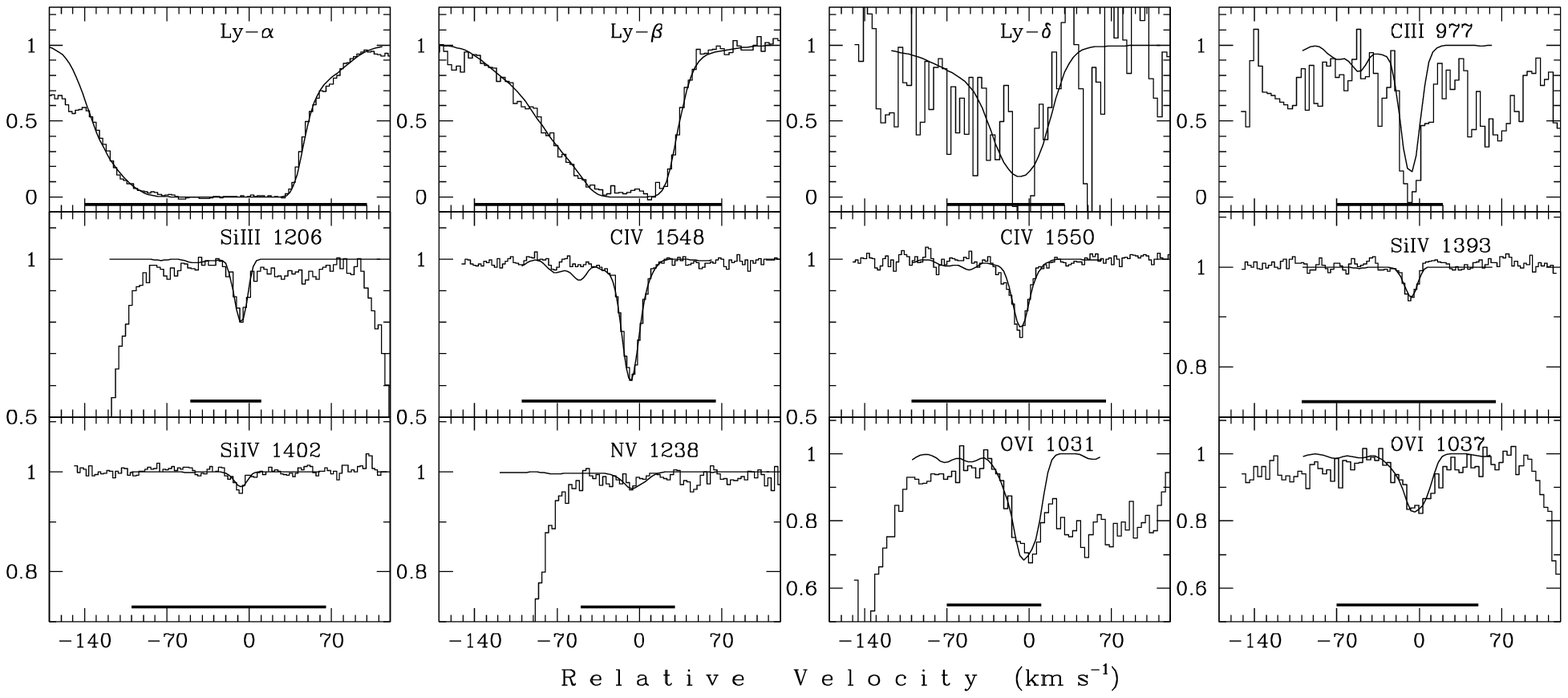,height=14.0cm,width=18cm}
\vspace{-8.0cm}
\caption[]{
Same as Fig.~\ref{fg_3} but for the \zabs = 2.379 system  towards
\object{HS 1700+6416}.  
Synthetic profiles (smooth curves) were calculated with the
recovered ionizing spectrum shown by the dotted line in Fig.~\ref{fg_13}.
The zero radial velocity is fixed at $z = 2.3799$. 
The estimated physical parameters 
are listed in Table~\ref{tb_3}, Col.~5.
}
\label{fg_16}
\end{figure*}

\subsubsection{Absorption system at \zabs = 2.379}

A set of metal lines observed in this system (Fig.~\ref{fg_16}) 
is identical to that from the \zabs = 2.568 absorber described
in Sect.~3.3.1.
Unfortunately, a noisy profile of \ion{C}{iii} hampers the accurate 
reconstruction of the spectral shape of the underlying radiation. 
Although this line lies deep in the Ly$\alpha$ forest,
it is very narrow ($FWHM = 22.5$ \kms) and cannot be a hydrogen line.
On the other hand, no metal candidate for possible blending was found.
Thus, the observed absorption is with high probability due to 
\ion{C}{iii}. 
However,  the spectrum of HM underestimates the observed intensity of 
\ion{C}{iii} far above the noise level, whereas the spectrum derived 
for the \zabs = 2.433 system underestimates \ion{Si}{iii}. 
At the same time, the spectral shape from \zabs = 2.568 system turned out 
to be quite consistent with all observed intensities.
The physical parameters obtained with this UVB are given in Table~\ref{tb_3}, 
Col.~5 with the corresponding synthetic profiles shown by the
solid lines in Fig.~\ref{fg_16}.

\begin{figure*}[t]
\vspace{0.0cm}
\hspace{0.1cm}\psfig{figure=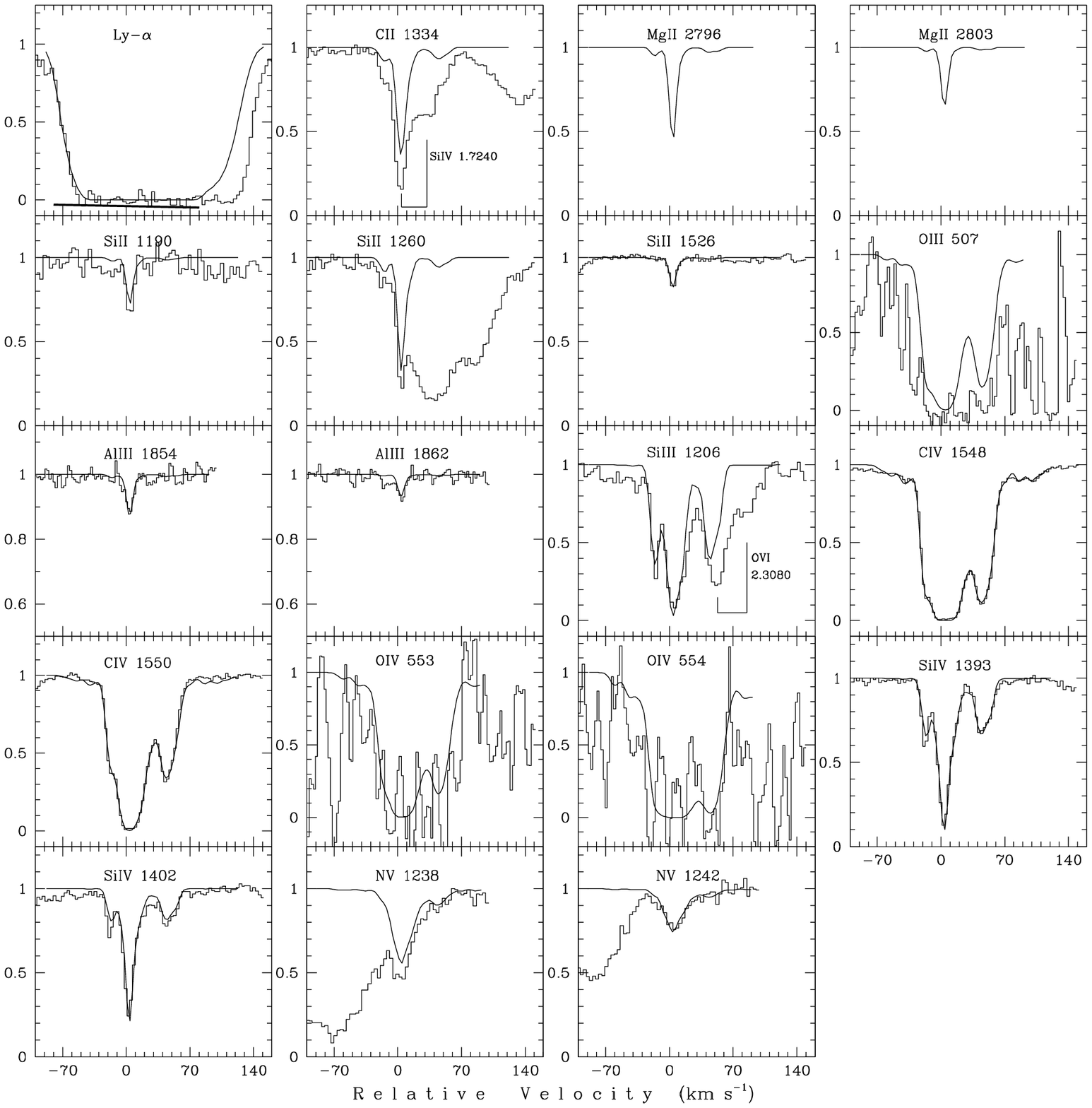,height=16.0cm,width=18cm}
\vspace{-0.5cm}
\caption[]{
Same as Fig.~\ref{fg_3} but for the \zabs = 1.845 system  towards
\object{HS 1700+6416}.  
Synthetic profiles (smooth curves) were calculated with the
recovered ionizing spectrum shown by the long-dashed line in Fig.~\ref{fg_13}.
The zero radial velocity is fixed at $z = 1.8450$. 
The estimated physical parameters 
are listed in Table~\ref{tb_3}, Col.~6.
Blends are indicated by tick marks.
}
\label{fg_17}
\end{figure*}

\subsubsection{Absorption system at \zabs = 1.845}

This system reveals one hydrogen line (Ly$\alpha$) and 
a lot of metals in different ionic transitions
(Fig.~\ref{fg_17}) which allow  us
to probe the underlying ionizing background. 
The column density of neutral hydrogen was estimated from low resolution spectrum 
by Vogel \& Reimers (1995) as 
5.6$\times10^{16}$ \cm , i.e. 
this system is optically thin in the hydrogen continuum. 
It is thin in the \ion{He}{ii} Lyman continuum as well:  
the attenuation of the incident radiation due to continuous
\ion{He}{ii} absorption in the cloud itself leads to an increasing 
\ion{Si}{iv}/\ion{C}{iv} ratio, but we find 
in this system \ion{Si}{iv}/\ion{C}{iv} = 0.088 which is
well below the values measured in the systems
with local attenuations at \zabs = 2.735, 2.944, and 2.875 (Tables~\ref{tb_1}
and \ref{tb_2}). 
Assuming optical depth in \ion{He}{ii} continuum below 
1 one obtains from $N$(\ion{He}{ii})~= $\eta\cdot N$(\ion{H}{i}) 
$ < 5\times10^{17}$ \cm the vlue of
$\eta \sim 10$ which assumes a
low ionization parameter (high gas density)
and a hard ionizing spectrum. 

A hard ionizing continuum is also required by the 
presence of \ion{N}{v}, otherwise we obtain [N/C]~$> 0$.  
Such overabundance of nitrogen is predicted at high over solar
metallicities by chemical evolution models 
(e.g. Hamann \& Ferland 1999),
but at $Z \sim Z_\odot$ still deficit of nitrogen, [N/C]~$< 0$, 
is measured (e.g. Srianand \& Petitjean 2001;
Reimers et al. 2005a; Jenkins et al. 2005). 
In principle, we cannot exclude a possibility that the present system
is out of equilibrium 
and \ion{N}{v} is still overionized 
(i.e. not in the ionization equilibrium with the background radiation). 
The \ion{O}{vi} lines could clear up the actual physical
conditions but they fall beyond the observable range. 
However, we do not observe any velocity shifts between the \ion{N}{v} 
lines and all other ionic transitions 
which could indicate a multiphase nature of the absorbing gas. 
This makes  `normal' (equilibrium) conditions preferable. 

Another very important clue to the physical properties of the absorbing gas 
deliver silicon ions: the observed ratios of 
\ion{Si}{iv}/\ion{Si}{iii}~$\simeq 1$ 
and 
\ion{Si}{iv}/\ion{Si}{ii}~$\simeq 10$ 
can be realized only at metallicities near solar values. 
As already explained in Sect.~3.1.2, this
stems from a strong dependence of silicon ion fractions 
on the temperature which at high metallicities
overrides dependence on the spectral shape of the UVB.

Taking together, these considerations allow us
to design a possible shape 
of the ionizing continuum. It is
shown in Fig.~\ref{fg_13} by
the long-dashed line with  
the derived physical parameters given 
in Table~\ref{tb_3}, Col.~6 and 
synthetic profiles plotted by the solid lines in Fig.~\ref{fg_17}.
The red wing of a sole hydrogen line is inconsistent with the assumption
of constant metallicity inside the absorber and we calculated it using the
density--velocity distribution obtained from metal lines.
As mentioned above, the available silicon lines require
metallicity close to solar. However, with $Z \simeq 0.5Z_\odot$ we
obtained $N$(\ion{H}{i})~$= 2.5\times10^{16}$ \cm\ which is already
half that 
estimated from the Lyman edge in low resolution spectrum.
A possible explanation of this discrepancy may be that we observe 
a high metallicity ($Z \sim Z_\odot$) blob with 
$N$(\ion{H}{i})~$\sim$ (1-2)$\times10^{16}$ \cm embedded in a cloud or
a chain of clouds with a joint column density 
$N$(\ion{H}{i})~$\sim$ (3-4)$\times10^{16}$ \cm.
To confirm this assumption, higher order \ion{H}{i} Lyman series lines
are needed.

Because of this uncertainty of $N$(\ion{H}{i}) in the metal-bearing cloud
and unknown element ratios, the spectral shape plotted in 
Fig.~\ref{fg_13} is not unique. However, it delivers quite reasonable
and self-consistent physical parameters and, thus, can be considered as
a plausible solution. In any case, the spectral indices 
$\alpha > -1.2$ in the range $1 < E < 4$ Ryd are favored since
for lower indices one would obtain C/Si above solar   
which has never been observed and is not predicted theoretically.

Taking together, the high metallicity of the blob and hard ionizing
spectrum allow to suggest the presence of a QSO/AGN close to the line of
sight.

\section{Discussion}
 
The spectral shape of the background ionizing radiation 
in the energy range $1 < E < 10$ Ryd is reconstructed from
optically-thin metal absorption-line systems observed in QSO 
spectra at $z \approx 1.80$ and $2.38 < z < 2.94$.
The spectral shape of the UVB does fluctuate, however these 
fluctuations are ruled by different sources 
at two considered redshift intervals.

In the range $2.38 < z < 2.94$, the spectral shape is mostly affected
by radiative transfer processes in the clumpy IGM involving
absorption in both the \ion{He}{ii} Ly-$\alpha$ and Lyman continuum.
We confirm our result from Paper I:
the UVB at $2.4 <$ \zabs $< 3.0$ demonstrates 
the intensity depletion between 3 and 4 Ryd 
caused by the absorption of the \ion{He}{ii} Ly$\alpha$ photons in 
the IGM (\ion{He}{ii} Gunn-Petersen effect). 
None of the recovered UVB shapes 
shows features which can be attributed
to the contribution of radiation from soft sources like 
starburst galaxies. Such
contribution would lead to a very steep slope at $E > 1$ Ryd 
($\alpha < -2.0$ at $1 < E < 3$ Ryd) and reduced
intensity at 3 Ryd (see Figs.~1 and 3 in Giroux \& Shull, 1997) whereas
the metal-line systems considered here 
require relatively hard
($\alpha > -1.6$) spectral shapes with pronounced
peaks at 3 Ryd combined with strong attenuation at $E > 4$ Ryd~--- features
unambiguously related to the absorption in the \ion{He}{ii} Lyman
continuum and subsequent re-emission of recombinant \ion{He}{ii}
Ly-$\alpha$.
 
\begin{figure}[t]
\vspace{0.0cm}
\hspace{-0.7cm}\psfig{figure=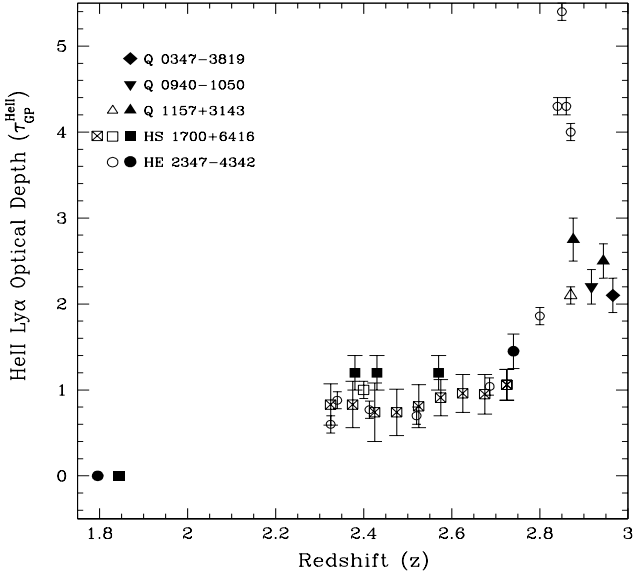,height=9cm,width=10cm}
\vspace{-1.3cm}
\caption[]{
Redshift dependence of the \ion{He}{ii} Ly$\alpha$ opacity,
$\tau^{\rm He\,II}_{\rm GP}(z)$.
Open symbols show direct measurements from
\object{HE 2347--4342} (Zheng et al. 2004), 
\object{HS 1700+6416} (open square~-- Davidsen et al. 1996,
crossed open square~-- Fechner et al. 2006a), and
\object{Q 1157+3143} (Reimers et al. 2005b). 
Filled symbols represent the values of 
$\tau^{\rm He\,II}_{\rm GP}$ 
estimated from the restored spectral shapes of the UVB ionizing radiation:
diamond and reverse triangle~--- Paper~I; triangles, squares and 
circles~--- the present paper. 
Error bars correspond to $1~\sigma$ confidence level.
}
\label{fg_18}
\end{figure}

From the depth of the GP trough in the recovered UVB spectra 
we can estimate the effective \ion{He}{ii}
Ly$\alpha$ opacity of the IGM, $\tau^{\rm He\,II}_{\rm GP}$. 
Fig.~\ref{fg_18} shows the obtained redshift evolution of
$\tau^{\rm He\,II}_{\rm GP}$ (filled symbols) together with 
the available directly measured \ion{He}{ii} Ly$\alpha$ opacities
(open symbols) along different sightlines. 
The concordance between the recovered values
and the direct measurements is good although our estimations are
systematically higher by 20-40\% ($\sim$ 1-2$\sigma$).  
This difference can be explained by the following reasons. 
Firstly, the approximation of the continuum depression between 3 and 4 Ryd 
by a straight step does not take into account a winding shape of
the actual GP trough. Secondly, the depth of this step is determined
mostly from the analysis of carbon lines \ion{C}{ii}, \ion{C}{iii}, and
\ion{C}{iv} (see Sect.~2). However,
the ionization potential of \ion{C}{iii}
(3.52 Ryd) is close to the onset of the \ion{He}{ii} Ly$\beta$ forest
(3.56 Ryd) and an
additional GP absorption in \ion{He}{ii} Ly$\beta$ 
can affect the fraction of this ion 
and the subsequent \ion{C}{iv}. 
These may lead to the overestimation of $\tau^{\rm He\,II}_{\rm GP}$
by 0.1--0.2 in our method.
On the other hand, estimations of
$\tau^{\rm He\,II}_{\rm GP}$ from FUSE data may give too small
values due to problems in background subtraction at short wavelengths 
corresponding to $z < 2.55$ (see Sect.~2 in F06).

Analysis of the close absorption systems (\zabs = 2.735 and 2.739 towards 
\object{HE 2347--4342}  
and \zabs = 2.433 and 2.438 towards \object{HS 1700+6416}) 
revealed two-three times fluctuations in $\eta$ between them. 
These systems share the same UVB, and the fluctuations in $\eta$
are caused  entirely by the 
variations in the absorber gas density.
 
Now compare the \zabs = 2.433 and 1.845 systems. Both of them show
similar $U$ and $n_{\rm H}$, but their $\eta$ values differ by
about 3 times (Table~3). In this case we observe the effect of 
different spectral softness (see Fig.~\ref{fg_13}).
 
\begin{figure}[t]
\vspace{0.0cm}
\hspace{-0.7cm}\psfig{figure=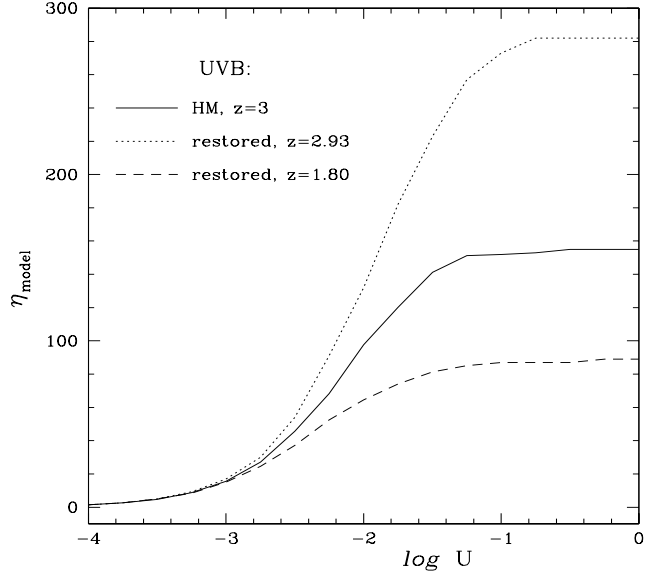,height=9cm,width=10cm}
\vspace{-1.3cm}
\caption[]{
Dependence of $\eta_{\rm model}$ 
on the ionization parameter $U$ for different spectral shapes
of the ionizing radiation:
HM at $z = 3$ (solid line), 
and UVB restored at $z = 2.93$ (dotted line)
and $z = 1.80$ (dashed line) which are, respectively, 
softer and harder than the HM spectrum.
The UVB at $z = 2.93$ (\object{Q 1157+3143}) is shown in Fig.~\ref{fg_9},
and  at $z = 1.80$ (\object{HS 0747+4259})~--- in Fig.~\ref{fg_20}.
The adopted metallicity is $Z = 0.01Z_\odot$.
}
\label{fg_19}
\end{figure}

Fig.~\ref{fg_19} shows the dependence of 
$\eta_{\rm model} = Z_{\rm He}\Upsilon_{\rm He\,II}/\Upsilon_{\rm H\,I}$    
on the ionization parameter $U$ calculated with
CLOUDY for optically thin case assuming different ionizing spectra. 
In case of a constant gas density inside the absorber, 
 $\eta_{\rm model} = \eta$. When the gas density fluctuates
within the absorber, $\eta$ is defined as a density weighted mean of
the $\eta(U)$ values, thus being lower than $\eta_{\rm model}$
corresponding to the mean ionization parameter $U_0$
(see eqs.[24,25] in LAK). 
It is seen that in the range $-3 < \log(U) < -1$ relevant for
optically thin
high redshift intergalactic absorbers the $\eta_{\rm model}$-curve demonstrates 
a large gradient which becomes
steeper with the softening of the UVB at $E > 4$ Ryd. 
Moreover, the interval of $\eta_{\rm model}$ values 
corresponding to $-3 < \log(U) < -1$ also
expands. Thus, in general, the scatter in the measured values of $\eta$ can 
arise from the combined action of 
differences in the mean gas density between the absorbers and
variable softness of the ionizing radiation due to absorption in
the \ion{He}{ii} Lyman continuum. 
We can also predict that the scatter in $\eta$ should decrease with
decreasing $z$ due to progressive hardening of the UVB spectrum.

Fig.~\ref{fg_19} explains also the correlation between 
low \ion{H}{i} column densities and high $\eta$ values:
weak absorbers tend to have lower gas densities and thus are 
shifted to the right part of the $\eta (U)$ curve. 
This is also one of the reasons why $\eta$ in voids are
systematically higher than in the filaments: by definition, 
voids consist of weak \ion{H}{i} absorbers.
Another reason is a possible shielding of the UV radiation in voids by 
the surrounding clouds which can 
soften the ionizing spectra. This is illustrated 
by the \zabs = 2.735, 2.739 and 2.741 absorbers towards 
\object{HE 2347--4342} located at the
boundary of the void at $2.718 < z < 2.727$ 
(Fig.~5 in Z04).
These absorbers belong probably to the filamentary structure and
form a layer with a linear size of $\sim 300$ kpc~\footnote{
It is interesting to note that other absorbers from our study with
characteristics similar to those of the complex at $z = 2.735-2.741$,
namely the \zabs = 2.917 system towards \object{HE 0940--1080} 
(Sect.~3.1, Paper I) and the \zabs = 2.875 system towards
\object{Q 1157+3143} (Sect.~3.2.3, present paper), also lie at the
boundary of a void.}. 
Table~2 from Z04 shows that
the $\eta$ values in the void at $2.718 < z < 2.727$ are significantly 
higher than in the void at $2.816 < z < 2.823$.
The latter has no strong \ion{H}{i} absorbers at its boundaries and 
lies close to the quasar \object{HE 2347--4342}, i.e.
the UVB in this void can be quite hard.
However, over large ranges in $z$ the correlation
``low $N$(\ion{H}{i})~--~high $\eta$'' 
[$N$(\ion{H}{i})~$= 10^{12}-10^{13}$ \cm] 
should be weak due to superposition of UVBs with
variable softness:
harder UVBs give a significantly lower $\eta$ especially for
low gas density (and, in general, low column density) absorbers,
thus effectively lowering the left side of the 
``$N$(\ion{H}{i})~--~$\eta$'' distribution. As for stronger
\ion{H}{i} absorbers   
[$N$(\ion{H}{i})~$= 10^{13}-10^{15}$ \cm],
they have usually higher gas densities (lower $U$ values) and
according to Fig.~\ref{fg_19} their $\eta$ values are less sensitive
to the variations of UVB. Thus, the correlation
``high $N$(\ion{H}{i})~--~low $\eta$'' should hold over all redshifts.
This is in concordance with direct measurements of $\eta$ from
the \ion{H}{i}/\ion{He}{ii} Ly$\alpha$ forest in F06
(see their Fig.~10) and Fechner \& Reimers (2006).

\begin{figure}[t]
\vspace{0.0cm}
\hspace{-0.2cm}\psfig{figure=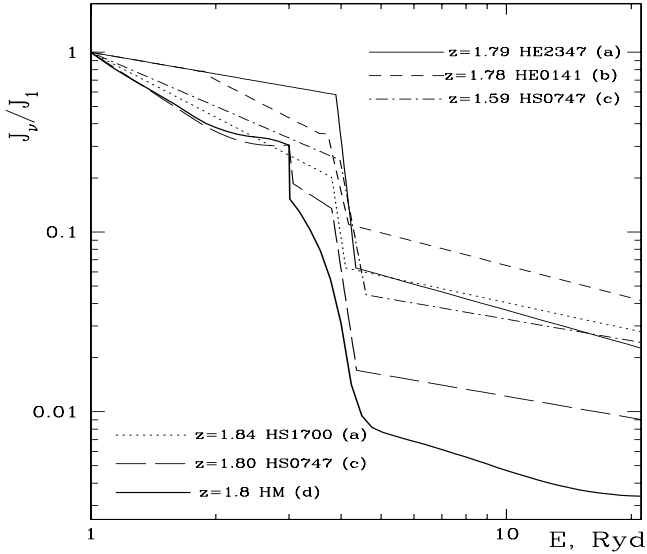,height=9cm,width=12cm}
\vspace{-1.3cm}
\caption[]{
The difference between
the UVB predicted by HM at $z = 1.8$ ({\it thick line})
and those restored by the MCISS procedure ({\it thin lines}).
The data are taken from 
(a)~--- the present paper, quasars \object{HE 2347--4342} and
\object{HS 1700+6416}, (b)~--- Reimers et al. (2005a),
quasar \object{HE 0141--3932}, (c)~--- Paper~II,
quasar \object{HS 0747+4259}, and (d)~--- Haardt \& Madau (1996).
}
\label{fg_20}
\end{figure}

As for UVBs recovered at 
$z \simeq 1.8$, they are very hard at $E > 3$ Ryd~--- much harder than
predicted by the HM model (Fig.~\ref{fg_20}). The reason for this
discrepancy was already given in our Paper II: 
HM used a biased statistics of the Ly$\alpha$ clouds
at \zabs $\la$ 1.8 with too many
strong Ly$\alpha$ absorbers leading to a significant underestimation of  
the UVB intensity at $E > 3$ Ryd.
From the depth of the \ion{He}{ii} break at 4 Ryd we can calculate the
effective \ion{He}{ii} column density which gives rise to this break 
(Table~4). It is seen that the values of $N$(\ion{He}{ii})
are spread by maximum 40\% around the median  
$1.2\times10^{18}$ \cm. 
Besides, the recovered ionizing spectra show
a diversity of shapes with a tendency to high spectral 
indices at $1 < E < 4$ Ryd ($\alpha > -1.4$, median $\alpha = -1.0$). 
These spectra are likely to be 
intrinsic spectra of quasars, i.e. not processed by the IGM.
In order to cause a noticeable absorption in the \ion{He}{ii} Lyman
continuum, the intervening absorbers should have 
$N$(\ion{He}{ii})~$ > 5\times10^{17}$ \cm. 
According to Fig.~\ref{fg_19}, 
the corresponding lower limit on $N$(\ion{H}{i}) will be
$3.0\times10^{15}$ \cm for the HM spectrum at $z = 3$ (middle curve)
and $1.8\times10^{15}$ \cm and $5.5\times10^{15}$ \cm for the
softer and harder spectra (upper and lower curves, respectively). 
From statistics of the Ly$\alpha$ forest (e.g., Kim et al. 2002)
it is well known that strong \ion{H}{i} absorbers 
with $N$(\ion{H}{i})~$> 10^{15}$ \cm become more rare
with decreasing redshift. This results in less absorption at
the \ion{He}{ii} ionization edge leading to a harder UVB.
To produce  $N$(\ion{He}{ii})~$ > 5\times10^{17}$ \cm,
harder spectra, in turn, require more stronger \ion{H}{i} absorbers 
which are even more rare. 
Thus, by $z \simeq 1.8$ a large part of the QSO/AGN
emitted radiation remains unprocessed.

In support to this
consideration the following observation-based
arguments can be given.
Using a combination of several hard X-ray surveys, Ueda et al. (2003)
investigated cosmological evolution of the AGN luminosity function.
They found that the comoving spatial density of most luminous AGNs
(quasars) peaks at $z \simeq 2$ and after that decreases rapidly, but
spatial density of less luminous but orders of magnitude
more numerous AGNs  continues to rise till
$z \la 1$. On the other hand, a FUSE survey of the AGNs ($z < 1$)
by Scott et al. (2004)
shows that lower luminosity AGNs tend to have harder spectral slopes 
(median $\alpha = -0.6$) in the
range 650--1000 \AA\, (0.9-1.4 Ryd). 

Thus, we can conclude that a significant hardening of the UVB spectrum
occurs between $z = 2.38$ and $z = 1.80$.  
According to Fig.~\ref{fg_19},
this hardening would result in decrease of  $\eta$.
The redshift evolution of
$\eta$ reported in Fechner \& Reimers (2006) 
allows to suggest that hardening starts
even earlier, somewhere at $z \sim 2.5$. 
At the moment we cannot trace the UVB evolution between $z = 2.5$ 
and $z = 1.80$
because of scarcity of our absorbers in this redshift interval.
This will be a topic of our future studies.

\section{Conclusions}

\begin{enumerate}
\item[1.] Using the MCISS method based on realistic models of line
formation in stochastic media and adequate computational technique,
we restored the spectral shapes of the underlying ionizing radiation
from optically thin metal-line systems.
The UVB spectral shape does fluctuate in the redshift range studied,
$1.8 < z < 2.94$.
However, the fluctuations at $z \la 1.8$ and 
$2.4 < z < 2.94$ are ruled by different physical reasons.
At $2.4 < z < 2.94$, the fluctuations arise due to radiative transfer
processes in the clumpy IGM, whereas at $z \la 1.8$ the IGM becomes
almost transparent both in the \ion{H}{i} and \ion{He}{ii} Lyman continua
and the variability of spectral shapes comes from diversity
of spectral indices of the intrinsic QSO/AGN spectra. 
\item[2.] At $2.4 < z < 2.94$, the recovered spectral shapes
show the intensity depression between 3 and 4 Ryd which is interpreted as
a manifestation of the \ion{He}{ii} Gunn-Peterson effect.
The values of the mean \ion{He}{ii} Ly$\alpha$ opacity
calculated from the depth of the recovered GP trough correspond well
(within 1-2$\sigma$) to 
$\tau^{\rm He\,II}_{\rm GP}$ directly measured from the
\ion{H}{i}/\ion{He}{ii} Ly$\alpha$ forest towards the quasars studied.
This concordance confirms the applicability of the MCISS procedure for
studying \ion{He}{ii} absorption in the IGM.
\item[3.] The scatter of $\eta$~= $N$(\ion{He}{ii})/$N$(\ion{H}{i}) and
anti-correlation between $N$(\ion{H}{i}) and $\eta$ can be
explained by the combined action of the spectral shape variability due
to radiative transfer in the clumpy IGM and differences in 
the mean gas densities in the absorbing clouds.
Progressive hardening of the recovered UVB towards lower $z$
explains the corresponding decrease of the median $\eta$-values
reported in F06 and Z04.
\item[4.] All recovered UVBs are produced by QSO/AGN radiation
processed by the clumpy IGM and do not show features which can be
attributed to the input of radiation from soft sources like
starburst galaxies. The metal systems considered in the present paper
and in Paper~I have overdensities 10--100 implying that radiative
transfer effects both in the cloud and in the IGM can be significant.
This means that for regions with smaller overdensities the UVB
should be harder in general. 
However, such regions (voids) can be bounded by filament structures
optically thin in \ion{H}{i} but thick in \ion{He}{ii}. In this case, 
the incident radiation is effectively shielded and the UVB in voids
becomes softer.

\end{enumerate}

\begin{acknowledgements}
The work of IDA and SAL is partly supported by
the RFBR grant No. 06-02-16489 and 
by the Federal Agency for Science and Innovations
(grant NSh 9879.2006.2).
CF is supported by the DFG under RE 353/49-1.
DT is supported in part by NSF grant AST 0507717 and NASA grants
NAG5-13113 and HST-AR-10688.01-A.
\end{acknowledgements}

\begin{table}[t]
\centering
\caption{\ion{He}{ii} column densities estimated 
from the ionizing spectra shown in Fig.~\ref{fg_20}
}
\label{tb_4}
\begin{tabular}{cccc}
\hline
\noalign{\smallskip}
{\footnotesize Quasar} & 
{\footnotesize \zabs} & 
{\footnotesize $\tau^{\rm He\,II}_{\rm c}$ } &
{\footnotesize $N$(\ion{He}{ii})} \\
 & & {\footnotesize (at 4 Ryd)} & {\footnotesize (\cm)}\\
\noalign{\smallskip}
\hline
\noalign{\smallskip}
\object{HE 0141--3932} & 1.78 & 1.2 & 0.7E18\\[-1pt]
\object{HS 0747+4259} & 1.80 & 2.0 & 1.3E18\\[-1pt]
                      & 1.59 & 1.8 & 1.2E18\\[-1pt]
\object{HS 1700+6416} & 1.84 & 1.2 & 0.7E18\\[-1pt]
\object{HE 2347--4342} & 1.79 & 2.2 & 1.4E18\\[-1pt]
\noalign{\smallskip}
\hline
\end{tabular}
\end{table}

\end{document}